\documentclass[aps,twocolumn,prx,superscriptaddress,showpacs,tightenlines]{revtex4-1}
\usepackage{amsmath}
\usepackage{graphicx}
\usepackage{epsfig}
\usepackage{amsfonts}
\usepackage{color}
\usepackage{epstopdf}

\begin{document}

\newcommand{\nn}{\nonumber}
\newcommand{\ms}[1]{\mbox{\scriptsize #1}}
\newcommand{\msi}[1]{\mbox{\scriptsize\textit{#1}}}
\newcommand{\dg}{^\dagger}
\newcommand{\smallfrac}[2]{\mbox{$\frac{#1}{#2}$}}
\newcommand{\Tr}{\text{Tr}}
\newcommand{\ket}[1]{|#1\rangle}
\newcommand{\bra}[1]{\langle#1|}
\bibliographystyle{apsrev}
\newcommand{\pfpx}[2]{\frac{\partial #1}{\partial #2}}
\newcommand{\dfdx}[2]{\frac{d #1}{d #2}}
\newcommand{\half}{\smallfrac{1}{2}}
\newcommand{\s}{{\mathcal S}}
\newcommand{\jord}{\color{red}}
\newcommand{\kurt}{\color{blue}}

\title{ Shortcuts to adiabaticity for open quantum systems and a mixed-state inverse engineering scheme}

\author{S. L. Wu }
\email{slwu@dlnu.edu.cn}
\affiliation{School of Physics and Materials Engineering,
Dalian Nationalities University, Dalian 116600 China}

\author{W. Ma }
\affiliation{School of Physics and Materials Engineering,
Dalian Nationalities University, Dalian 116600 China}

\author{X. L. Huang}
\affiliation{School of Physics and Electronic Technology,
Liaoning Normal University, Dalian 116029, China}

\author{Xuexi Yi }
\email{yixx@nenu.edu.cn}
\affiliation{Center for Quantum Sciences and School of Physics,
Northeast Normal University, Changchun 130024, China}

\date{\today}

\begin{abstract}
We propose a fast mixed-state control scheme to transfer the quantum state along designable trajectories in Hilbert space, which
is robust to multiple decoherence noises. Starting with the dynamical invariants of open quantum systems, we present the
shortcuts to adiabaticity (STAs) of open quantum systems at first, then apply the STAs to speed up the adiabatic steady process. Our
scheme drives open systems from a initial steady state to a target steady state by a controlled  Liouvillian that possesses the same form
as the reference  (original) one which is accessible in present-day experiments. The experimental observation with current available
parameters for the  nitrogen-vacancy (NV) center in diamond is suggested and discussed.
\end{abstract}

\maketitle

\section{Introduction}

Controlling quantum systems to accomplish special task is at the heart of emerging quantum technologies\cite{Koch2019,
Osnaghi2001,Hacohen2018}. The ideal control scheme needs to satisfy three important issues:  (i) High-speed: The quantum
state  transfers to the target state within desired control time length;  (ii) High-fidelity: The control process should be with
an admissible error; (iii) High-controllability: The trajectory from an initial state to a target state must be completely controllable.

The most selected schemes in experiment are based on the unitary evolution of  closed systems\cite{Thanopulos2007,Berry2009,
Chen2010,Chen2016}. Yet unwanted couplings to the environment reduce the fidelity severely\cite{Huneke2013, Zhou2017},
and the final state can not be steadied on the target state after the control is done. This shackles the quantum sciences and technologies
to realize efficient and scalable devices beyond the current circumstances of proof-of-principle demonstrations. To overcome such
shackles, a straightforward thinking is to formulate a scheme based on the theory of open quantum systems. Due to its steadiness, the
steady state becomes an important candidate to achieve the quantum control task\cite{Sarandy2005}. The adiabatic steady state
engineering scheme transfers the quantum state into the target state along instantaneous steady state\cite{Lorenzo2016,Wu2017}.
But the control process needs to be very slow so that the adiabatic condition can be satisfied. Also, the fast steady state engineering
schemes are proposed to accelerate the adiabatic evolution of the  equilibrium state of quantum thermodynamic system, which can only
be applied to the Gibbs states or the Gaussian states \cite{Dupays2020,Dann2019}. On the other hand, the transitionless quantum driving
method of open quantum systems is also proposed, but a feasible control protocol is hard to be presented\cite{Vacanti2014,Wu2019}.
Therefore, up to present, none of schemes satisfies the requirements including high-accuracy, high-controllability, and  high-speed at the
same time.

In this paper, we propose a fast control scheme of open quantum systems, named the mixed-state inverse engineering scheme (MIE),  which
allows a robust and precise transfer to a given target state with a designable mixed state trajectory.  Firstly, by analysing the spectral features
of dynamical invariant superoperators of open quantum systems\cite{Sarandy2007}, we present general solutions of quantum states governed
by the master equation
\begin{eqnarray}
\partial_t |\rho(t)\rangle\rangle=\hat{\mathcal L}_c (t)|\rho(t)\rangle\rangle.\label{deq1}
\end{eqnarray}
with the Liouvillian superoperator $\hat{\mathcal L}_c (t)$. Based on this general solution, the STAs of open quantum systems are established.
Then, we apply the STAs of open quantum systems on the steady state engineering (i.e., the MIE scheme), and show that the target state can be
reached with extremely high fidelity and within desired control time length. The central advantage of our scheme is that the control tasks are, in
general, achieved with  fruitful feastible control protocols in experiment by selecting different quantum state trajectories. More importantly,  the
selectable trajectories can bring the ideal final fidelity  even multiple noise sources  are involved. In fact, the pure-state inverse engineering scheme
of closed quantum systems is a particular case of the STAs scheme of open quantum systems \cite{Chen2011}, when the trajectories of quantum
states are the  pure-state trajectories. And the MIE scheme can overcome the difficulties in the earlier STAs methods for closed system due to
the designable mixed-state trajectory..

The rest of this paper is organized as follows. In Sec.\ref{method}, we present the STAs scheme of open quantum systems, and propose the
MIE scheme to accelerate the adiabatic steady state process. In Sec. \ref{stirap}, we apply the MIE scheme to NV center system, which provides
simple, practical, robust control protocols to transfer the population from one ground state to the other\cite{Chen2012,Baksic2016}. It is shown
that  the MIE scheme is far better at transfer efficiency than any STAs scheme of closed systems\cite{Zhou2017}, especially in the case that
the three level system suffers from dissipation and dephasing at the same time. In Sec.\ref{comparison}, we discuss  the relationship between  the
mixed-state inverse engineering scheme and the pure-state inverse engineering scheme of closed systems\cite{Chen2011}.  Conclusions
are presented in Sec. \ref{conclusion}.

\section{ Methods} \label{method}

\subsection{STAs of Open Quantum Systems}

For an open quantum system governed by Eq.(\ref{deq1}), a dynamical invariant of the open quantum system is defined as a superoperator
$\hat{\mathcal I} (t) $ which satisfies\cite{Sarandy2007}
\begin{eqnarray}
\partial_t \hat{\mathcal I} (t)- [\hat{\mathcal L}_c  (t),\hat{\mathcal I} (t)] =0.\label{di}
\end{eqnarray}
In general, invariants are non-Hermitian. Thus $\hat{\mathcal I} (t) $ needs to be expressed as the Jordan canonical form. Consider that there are $m$ Jordan
blocks, and the $\alpha$-th Jordan block is $n_\alpha$-dimensional. According to the Jordan decomposition of $\hat{\mathcal I} (t)$, we introduce
right vectors $\{|D_\alpha^{(i)}\rangle\rangle\}$ and left vectors $\{\langle\langle E_\alpha^{(i)}|\}$ in the Hilbert-Schmidt space. The left and right
vectors always satisfy
\begin{eqnarray*}
\hat {\mathcal I}\,|D_\alpha^{(i)}\rangle\rangle=\lambda_\alpha|D_\alpha^{(i)}\rangle\rangle+|D_\alpha^{(i-1)}\rangle\rangle,\\
\langle\langle E_\alpha^{(i)}| \hat {\mathcal I}=\lambda_\alpha\langle\langle E_\alpha^{(i)}|+\langle\langle E_\alpha^{(i+1)}|,
\end{eqnarray*}
with $|D_\alpha^{(-1)}\rangle\rangle\equiv0,\,\langle\langle E_\alpha^{(n_\alpha)}|\equiv0$ for $i=0,1,...,n_\alpha-1$. Thus, the right and left vectors
$|D_\alpha^{(0)}\rangle\rangle$ and $\langle\langle E_\alpha^{(n_\alpha-1)}|$ are the right and left eigenstates of $\hat {\mathcal I}(t)$ with the eigenvalue
$\lambda_\alpha$.  Here we assume that all of eigenvalues  are nondegenerate, i.e., $\lambda_\alpha\neq\lambda_\beta$ for $\forall\, \alpha\neq\beta$.
And the left and right vectors  satisfy the orthonormality condition
\begin{eqnarray*}
\langle\langle E_\alpha^{(i)}|D_\beta^{(j)}\rangle\rangle=\delta_{\alpha\beta}\delta_{ij}
\end{eqnarray*}
It can be verified that the eigenvalues of the dynamical invariants are time-independent, and
\begin{eqnarray*}
\langle\langle E_\beta^{(j)}|\hat O|D_\alpha^{(i)}\rangle\rangle=0,\,\,\forall\,i,\,j,\label{a12}
\end{eqnarray*}
with $ \hat O=\hat{\mathcal L}-\partial_t$ for $\alpha\neq\beta$. Therefore, it can be verified that the general solution of Eq.(\ref{deq1}) reads
\begin{eqnarray}
|\rho(t)\rangle\rangle=\sum_{\alpha=0}^{m-1} c_\alpha\,\exp(\eta_\alpha(t))\,|\Phi_\alpha(t)\rangle\rangle,\label{fs1m}
\end{eqnarray}
 in which $\eta_\alpha(t)$ is a complex phase, $c_\alpha$ is a time-independent expansion efficient. $|\Phi_\alpha(t)\rangle\rangle$ is a  right vector
in  the $\alpha$-th Jordan block, which can be written as
\begin{eqnarray}
|\Phi_\alpha(t)\rangle\rangle=\sum_{i=0}^{n_\alpha-1} b^\alpha_i(t)|D_\alpha^{(i)}(t)\rangle\rangle,
\end{eqnarray}
with coefficients $b^\alpha_i(t)$. The details of the derivation of the general solution Eq.(\ref{fs1m}) can be found in Appendix. \ref{AA}.

Since the adiabaticity of open quantum systems requires only forbidding the transition between different Jordan blocks\cite{Sarandy2005}, the general solution
Eq.(\ref{fs1m}) is enough to establish STAs of open quantum systems. Suppose that our aim is to drive the quantum system from an initial Liouvillian $\hat
{\mathcal L}_c(0)$ to a final one $\hat {\mathcal L}_c(t_f)$, such that the ``population" in the initial and final instantaneous Jordan blocks are same but admitting
transitions at the intermediate times. Based on the general solution Eq.(\ref{fs1m}), the complex phases $\eta_\alpha(t)$ are chosen as arbitrary functions to write
down the time-evolution superoperator $\hat {\mathcal E }(t)$ as
\begin{eqnarray*}
\hat {\mathcal E}(t)=\sum_{\alpha=0}^{m-1} \exp\left(\eta_\alpha(t)\right)|\Phi_\alpha(t)\rangle\rangle \langle\langle\Psi_\alpha(0)|,
\end{eqnarray*}
 where $\langle\langle\Psi_\alpha(t)|$ is  a left vector of the $\alpha$-th Jordan block, which satisfies
$\langle\langle\Psi_\beta(t)|\Phi_\alpha(t)\rangle\rangle=\delta_{\alpha\beta}$.  The evolution superoperator obeys
\begin{eqnarray*}
\partial_t \hat{ \mathcal E}(t)=\hat{\mathcal L}_c (t)\hat{\mathcal E}(t),
\end{eqnarray*}
which we formally solve for the control Liouvillian
\begin{eqnarray*}
\hat{\mathcal L}_c (t)=\partial_t \hat{ \mathcal E}(t)\hat{\mathcal E}^{-1}(t),
\end{eqnarray*}
with $$\hat{\mathcal E}^{-1}(t)=\sum_{\alpha=0}^{m-1} \exp\left(-\eta_\alpha(t)\right)|\Phi_\alpha(0)\rangle\rangle \langle\langle\Psi_\alpha(t)|.$$
Thus, we can express the control Liouvillian superoperator as
\begin{eqnarray}
&&\hat{\mathcal L}_c (t)=\nonumber\\
&&\sum_{\alpha=0}^{m-1} \left(|\partial_t \Phi_\alpha(t)\rangle\rangle \langle\langle\Psi_\alpha(t)|+\partial_t\eta_\alpha(t)
|\Phi_\alpha(t)\rangle\rangle \langle\langle\Psi_\alpha(t)|\right).\label{contliou}
\end{eqnarray}
Note that for a given dynamical invariant, there are many possible Liouvillians corresponding to different choices of complex phases $\eta_\alpha(t)$.
In general, $\hat{\mathcal I}(0)$ does not commute with $\hat{\mathcal L}_c (0)$, which implies that the Jordan blocks of
$\hat{\mathcal I}(0)$ do not coincide with the Jordan blocks of $\hat{\mathcal L}_c (0)$. $\hat{\mathcal L}_c (t_f)$ does not necessarily commute with
$\hat{\mathcal I}(t_f)$ either. We impose $[\hat{\mathcal I}(0),\hat{\mathcal L}_c (0)]=[\hat{\mathcal I}(t_f),\hat{\mathcal L}_c (t_f)]=0$, such that
the Jordan blocks coincide and then the quantum state transfer from the initial block to the final one is guaranteed. Here, we must emphasize
that $|\Phi_\alpha(t)\rangle\rangle$ can be arbitrary superposition of the right basis vectors $\{|D_\alpha^{(i)}(t)\rangle\rangle\}_{i=0}^{n_\alpha-1}$
for the $\alpha$-th Jordan block. In other words, if the quantum state is prepared in a given Jordan block of $ \hat{\mathcal L}_c (0)$ at the beginning
and the final state is still in the same block of $\hat{\mathcal L}_c (t_f)$, the shortcuts to adiabaticity of open quantum systems is established, which is
the control Liouvillian given by Eq.(\ref{contliou}) designed to. This is the first result of this paper.

In Appendix. \ref{AB}, we present a detailed comparison between the STAs scheme and the transitionless quantum driving scheme of open quantum
systems\cite{Vacanti2014}.  It is shown that, if the trajectory of the STAs scheme of open quantum systems is chosen
as the adiabatic trajectory, the STAs scheme is coincident with the transitionless quantum driving method proposed in Ref.\cite{Vacanti2014}. However, the
 adiabatic trajectory is not the only choice of the trajectories in our scheme. There are many trajectories can be used to inversely engineer the open quantum
 system. Proper trajectories  always provide reasonable and applicable control protocols, which helps us to overcome the difficulties met in the
control of   microscopic or/and mesoscopic systems.

\subsection{Mixed-state inverse engineering}

In a practical application, the general control Liouvillian presented in  Eq.(\ref{contliou}) will meets difficulties in giving practical and affirmatory
control protocols. In addition, most of  eigenvectors of the control Liouvllian  $ \hat{\mathcal L}_c(t)$ are unphysical quantum states, except the eigenvectors
with zero eigenvalues, which corresponds to steady states of open quantum systems. Therefore, for practical applications, we  focus our attention on the steady
states engineering of open quantum systems. Even if we  restricts our discussion on speeding up the adiabatic steady state process, it is
still difficult to obtain feasible control protocols  \cite{Wu2019}, since the control Liouvillian Eq.(\ref{contliou}) is in form of the superoperator.  In the following,  we
propose an effective and practical method to obtain feasible control protocols, which is easy to be used in experiment.

We consider  a quantum system with  $N$-dimensional Hilbert space govern  by a linear, time-local
master equation
\begin{eqnarray*}
\partial _t \rho(t)&=&\hat{ \mathcal L_0} (t)\mathcal [ \rho(t)]\nonumber\\
&=& -\frac{i}{\hbar}[{H}_0(t),{\rho}]+\sum_\alpha\hat {\mathcal D}[L_\alpha](\rho),
\end{eqnarray*}
where $\hat{\mathcal L_0}(t)$ is the reference Liouvillian  in the Lindblad form, $H_0(t)$ is the
Hamiltonian, and
\begin{eqnarray}
\hat{\mathcal D}[L_\alpha](\rho)=L_{\alpha}(N_\alpha){\rho}L_{\alpha}^{\dagger}(N_\alpha)-\frac{1}{2}\{L_{\alpha}^{\dagger}(N_\alpha)
L_{\alpha}(N_\alpha),{\rho}\}\nonumber\\\label{lindblad}
\end{eqnarray}
is the Lindbladian. The Lindblad operators $L_\alpha(N_\alpha)$ are related to some parameters $\{N_\alpha\}$, such as
the decoherence rates and the temperatures of the environments. Here, we do not limit the master equation to be
Markovian, but the corresponding evolution  must be a completely positive trace-preserving map.  Further, we assume that
$\hat{\mathcal L_0}(t)$ admits an unique (instantaneous) steady state $\rho_0(t)$, which satisfies $$\hat{\mathcal L_0}(t)[\rho_0(t)]=0.$$

For practical applications, we focus our attention on speeding up the adiabatic steady state process\cite{Lorenzo2016}, and seek the control
Liouvillian from Eq.(\ref{di}) directly. Concretely, the  control task is to drive the open  quantum system from the steady state of an initial Liouvillian
$\hat{\mathcal L_0} (0)$ to the target one $\hat{\mathcal L_0} (t_f)$\cite{Lorenzo2016,Sarandy2005}.  In order to achieve such purpose, we consider
that the invariant has only one 1-dimensional Jordan block with non-zero eigenvalue $\Omega_\text{I}$. If we set $| \Phi_0 (0)\rangle\rangle=
|\rho_0(0)\rangle\rangle$ and $| \Phi_0 (t_f)\rangle\rangle=|\rho_0(t_f)\rangle\rangle$, the eigenvector $|\Phi_0 (t)\rangle\rangle$ corresponds
to the trajectory connecting the initial steady state and the target steady state.   In general, we need to parameterize  $\hat{\mathcal I} (t) $ in
$N^2$-dimensional Hilbert-Schmidt space with $N^4$ independent coefficients \cite{Wu2019}. For the MIE scheme, since $| \Phi_0(t)\rangle\rangle$
should be a quantum  state of the open quantum system, we can expand $| \Phi_0(t)\rangle\rangle$ by right vectors $\{|T_\mu\rangle\rangle\}
_{\mu=1}^{N^2-1}$ which correspond to the SU($N$)  Hermitian generators $\{T_\mu\}_{\mu=1}^{N^2-1}$ , i.e.,
\begin{eqnarray}
| \Phi_0 (t)\rangle\rangle=\frac{1}{N}\left ( | \text I\rangle\rangle+\sqrt{\frac{N (N-1)}{2}}\sum_{\mu=1}^{N^2-1} r_\mu |T_\mu\rangle\rangle\right ),
\end{eqnarray}
where $\vec r= (r_1, r_2, ..., r_{N^2-1}) $ is the generalized Bloch vector with $\sum_\mu |r_\mu|^2<1$, and $|\text{I}\rangle\rangle$ is the  right vector
corresponding to a $N\times N$ identical operator. Thus, the dynamical invariants used in the MIE scheme can be defined as
\begin{eqnarray}
\hat{\mathcal{I}}(t)=\Omega_\text{I}|\Phi_0(t)\rangle\rangle\langle\langle \text{I}|,\label{di11}
\end{eqnarray}
where $\Omega_\text{I}$ is an arbitrary nonzero constant and $\langle\langle \text{I}|$ is the left vector corresponding to a $N\times N$ identity matrix.
With this notation, we can parameterize the dynamical invariant $\hat{\mathcal I}(t)$ with only $N^2-1$
parameters, which greatly simplifies the procedure in formulating control protocols.

To formulate feasible control protocols, we impose the control Liouvillians take the form as
\begin{eqnarray}
\hat{ \mathcal L}_c[\bullet]=-\frac{i}{\hbar}[{H}(t),{\bullet}]+\sum_\alpha\hat {\mathcal D}[L_\alpha](\bullet),
\end{eqnarray}
 in which the Lindbladians and the Hamiltonian are chosen according to the following principles:  (i) The Lindbladians in the control Liouvillian
$\hat{\mathcal L}_c (t) $ have  the same form as Eq.(\ref{lindblad}), which controls the system through control parameters  $\{N_\alpha\}$.
(ii) The control can also exert on the system via the Hamiltonian in $\hat{\mathcal L}_c(t).$  As it can be written
in terms of  SU (N)  Hermitian generators  $\{T_k\}$, i.e., $$H (t) =\sum_{k=1}^{N^2-1}c_k(t)T_k,$$ we might choose $\{c_k (t)\}$ as the
control parameters to manipulate the system. In this way, the control Liouvillians can always present feastible control protocols.
Substituting $\hat{\mathcal L}_c(t)$ and $\hat{\mathcal I} (t) $ into Eq.(\ref{di}), we can express the control parameters $\{c_k, N_\alpha\}$
as a function of the generalized Bloch vector and its  derivative $\{r_\mu, \partial_t r_\mu\}$.  On the other hand, the control Liouvillian is
always the same as the reference Liouvillian at the initial and final moment, which leads to boundary conditions for $\{r_\mu,\partial_t r_\mu\}$.
By utilizing the control parameters $\{c_k, N_\alpha\}$ and setting proper boundary conditions for $\{r_\mu, \partial_t r_\mu\}$, the
open quantum system can be transferred from the initial steady state into the target steady state along an exact trajectory given by $|\Phi_0 (t)
\rangle\rangle$. This is the second result of this paper.

If some  control parameters, say $\{\tilde c_k, \tilde N_\alpha\}$, are difficult to implement   in a real setting, we can single out the equations for
those parameters, and force them to be some values which are available in  the experimental setting. Notice that the equations for $\{\tilde c_k,
\tilde N_\alpha\}$ are about the components of the trajectory $\{r_\mu, \partial_t r_\mu\}$. Thus picking up proper $\{\tilde r_\mu\}$ to be
 free components in trajectory, we obtain a set of
 differential equations about $\{\tilde r_\mu\}$. Solving those differential equations is equivalent to choose a trajectory with particular components
$\{\tilde r_\mu\}$. Therefore the MIE scheme can avoid those difficulties encountered  in the control process, such as the negative decoherence
rate\cite{Alipour2020} and the impractical energy-levels couplings\cite{Chen2010}.  As a result,  the MIE offers  a practical
method to engineer an open quantum system with an exact trajectory which can be realized in  laboratory with current technology.

\section{Example:  The stimulated Raman adiabatic passage (STIRAP)}\label{stirap}

\begin{figure}[htbp]
\centerline{\includegraphics[width=0.9\columnwidth]{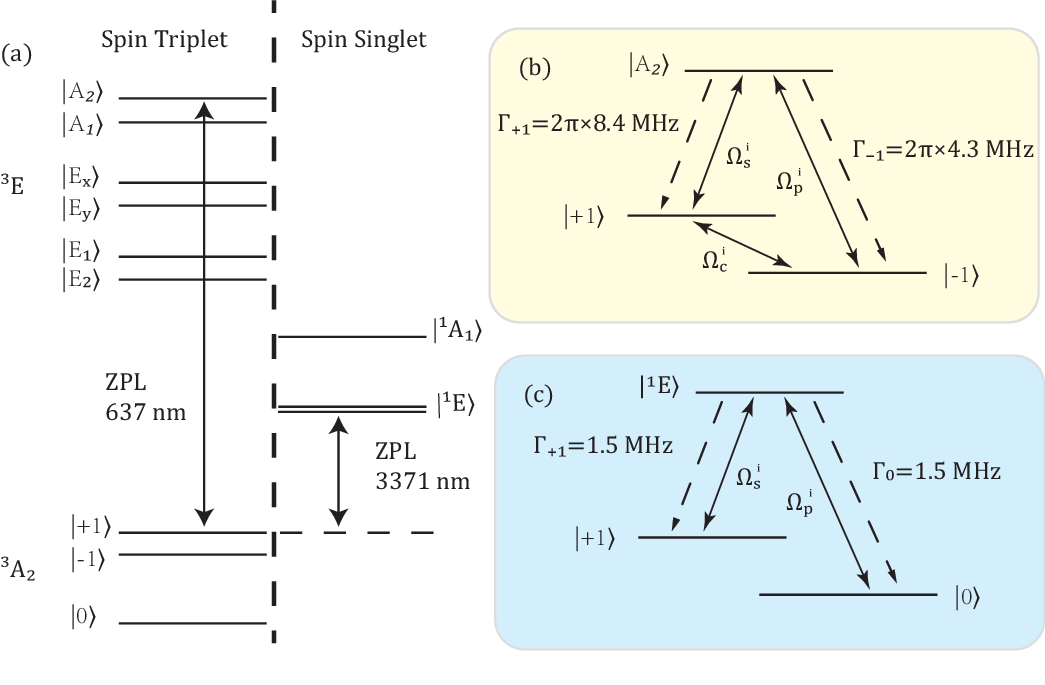}}
\caption{The level structure of the NV center. (a) A schematic illustration of the level structure of the NV centers.
The optical zero photon line (ZPL) at 637 nm is related to the transition from $^3E$ to  $^3 A_2$ and
the ZPL at 3371 nm corresponds to the transition from  $^1E$ to  $^3 A_2$.
(b) State transfer in a NV centre $\Lambda$ system by the protocol with initial-to-final state coupling. (c)
State transfer in a NV centre $\Lambda$ system by the protocol without the initial-to-final state coupling.}\label{il3}
\end{figure}

Consider a single nitrogen-vacancy (NV) center in diamond, which hosts a solid-state $\Lambda$ system.
The NV center has a spin-triplet, orbital-singlet ground state ($^3 A_2$) that is coupled optically to a spin-triplet,
orbital-doublet excited state ($^3 E$), as shown in Fig. \ref{il3} (a). The experiments have  had identified the three singlet
states ($\ket{^1E}$, $\ket{^1A_1}$) \cite{Rogers2008}, where $\ket{^1E}$ is double degenerate.

For the negative charged NV center with electron spin $S=1$, the ground state is a spin-triplet state with a zero-field
splitting $D_0=2.87$ GHz between spin sublevels $\ket{m_s=0}$ and $\ket{m_s=\pm 1}$ due to electronic spin-spin interaction.
Applying a static magnetic field $B_{NV}$ along the NV axis splits the $\ket{m_s=-1}$ and $\ket{m_s=+1}$
ground states by $2\gamma_{NV}B_{NV}$ with $\gamma_{NV}=2.8\,\text{MHz G}^{-1}$.
Passing single tunable laser (637.2 nm) through a phase electro-optic modulator
produce frequency harmonics to resonantly excite both $\ket{m_s=-1}$ and $\ket{m_s=+1}$ to the spin-orbit excited
state $\ket{A_2}$, which is used as the intermediate state for STIRAP. After the modulation of an amplitude electro-optic
modulator with a 10 GHz arbitrary wave form generator produces the control fields  $\Omega_s(t)$
and $\Omega_p(t)$ used in STIRAP. The Hamiltonian within the rotating wave approximation can be expressed in
the basis $\{\ket{m_s=-1},\ket{A_2},\ket{m_s=+1}\}$ by a matrix\cite{Carroll1988},
\begin{equation}
H_0(t)=\frac{\hbar}{2}
\left(
\begin{array}{ccc}
    0& \Omega_p(t)& 0\\
    \Omega_p(t)& 0& \Omega_s(t)\\
    0& \Omega_s(t)& 0\\
\end{array}
\right).\label{rh3}
\end{equation}
For simplify our discussion, the ``one-photon resonance'' case is considered. The shortcuts of the open STIRAP for a general case can
be obtained with the same procedure. Here we assume that the adiabatic pulses satisfy
\begin{equation}
\Omega_s(t)=\Omega(t)\cos\theta(t), \,\Omega_p(t)=\Omega(t)\sin\theta(t), \label{ap3}
\end{equation}
with $\tan\theta(t)=\Omega_p(t)/\Omega_s(t)$ and $\Omega(t)=\sqrt{\Omega_s^2(t)+\Omega_p^2(t)}$.

Consider that the $\Lambda$ system couples to a bosonic heat reservoir at finite temperature $T$.
The effect of the heat reservoir is to induce decay from $\ket{A_2}$ to $\ket{m_s=\pm1}$. The decay
rate are $\Gamma_{-1}=2\pi\times4.3$ MHz (from $\ket{A_2}$ to $\ket{m_s=-1}$) and $\Gamma_{+1}=2\pi\times8.5$ MHz  (from $\ket{A_2}$
to $\ket{m_s=+1}$), as shown in FIG. \ref{il3} (b). Moreover, the orbital dephasing of the level $\ket{A_2}$ can not
be ignored \cite{Issoufa2014}, where the dephasing rate is $\Gamma_d=2\pi\times8.8$ MHz.
All of decoherence rates mentioned here are selected from the measurement in recent experiment\cite{Zhou2017}. The dynamics of the $\Lambda$
system is governed by
\begin{equation}
\partial _t \rho(t)=\hat{ \mathcal L_0}\rho(t)+\Gamma_d\hat{ \mathcal D}[L_d]\rho(t),\label{me3}
\end{equation}
where
 \begin{eqnarray*}
\hat{ \mathcal L_0}\rho(t)&=&-\frac{i}{\hbar}[{H}_0(t),{\rho}]\\&+&\sum_{\alpha=0,+1}\Gamma_\alpha((N_\alpha +1)\hat {\mathcal D}[L_\alpha](\rho)
+N_\alpha\hat {\mathcal D}[L_\alpha^\dagger](\rho)). \nonumber
 \end{eqnarray*}
and $\hat{\mathcal D}[L_\alpha](\rho)=
L_{\alpha}(t){\rho}L_{\alpha}^{\dagger}(t)-\frac{1}{2}\{L_{\alpha}^{\dagger}(t)L_{\alpha}(t),{\rho}\}$. For the decay form  $\ket{A_2}$,
the lindblad operators can be expressed as $L_{\pm1}=\ket{m_s=\pm1}\bra{A_2}$; and $L_d=\ket{A_2}\bra{A_2}$ for the orbital dephasing.
$N_\alpha=[\exp(\hbar \omega_{2\rightarrow\alpha}/kT)-1]^{-1}$ denote the mean excitation numbers. In the following discussion, we choose
$\hat{ \mathcal L_0}$ as the reference Liouvillian, and  the  dephasing is the key obstacle for the performance of the protocol.

\subsection{The Adiabatic Trajectory}

In this subsection, we present the control protocol where the quantum state transfers along the adiabatic trajectory given by the instantaneous
steady state of $\hat{ \mathcal L}_0(t)$. We parameterize the instantaneous steady state of  $\hat{ \mathcal L}_0(t)$ via the generalized Bloch
vector $\{r_k\}_{k=1}^8$. The density matrix of the three-level system can be written as,
\begin{equation}
\rho(t)=\frac{1}{3}\left(\text{I}+\sqrt{3}\sum_{k=1}^8 r_k(t) T_k\right),\label{is3}
\end{equation}
where $\text{I}$ is a $3\times 3$ identity matrix, and $T_k$ denotes the regular Gellmann matrix. These $\{T_k\}$ span all traceless Hermitian
matrices of the Lie algebra su(3). If $\Gamma_{+1}=\Gamma_{-1}\equiv\Gamma$ and $N_{+1}=N_{-1}\equiv N$, the components of the Bloch
vector corresponding to  the instantaneous steady state of  $\hat{ \mathcal L}_0(t)$  are
\begin{eqnarray}
&&r_2=\sqrt{3}\,\mathrm{N}\, \mathrm{\Gamma}\, \mathrm{\Omega_p}/z,\nonumber\\
&&r_3=\sqrt{3}\,\left(\left(3\, \mathrm{N}^2 + 2\mathrm{N}\right)\, {\mathrm{\Gamma}}^2 +  {\mathrm{\Omega_s}}^2 \right)/(2z),\nonumber\\
&&r_4=-\sqrt{3}\,\mathrm{\Omega_p}\, \mathrm{\Omega_s}/z,\nonumber\\
&&r_7=-\sqrt{3}\,\mathrm{N}\, \mathrm{\Gamma}\, \mathrm{\Omega_s}/z,\nonumber\\
&&r_8=-\left(\left(3\, \mathrm{N}^2 + 2\mathrm{N}\right)\, {\mathrm{\Gamma}}^2+2\,{\mathrm{\Omega_p}}^2 -{\mathrm{\Omega_s}}^2 \right)
/(2z),\label{r38m}
\end{eqnarray}
with $z={\left(3\, \mathrm{N} + 1\right)\, {\mathrm{\Omega}}^2 + \mathrm{N}\, {\mathrm{\Gamma}}^2\, {\left(3\, \mathrm{N} + 2\right)}^2}$,
and the other components are zeros. The details for obtaining the instantaneous steady state can be found in Appendix. \ref{AC}.  Correspondingly,
 the dynamical invariants can be expressed by the Bloch vector according to the MIE scheme (see Eq.(\ref{di11})),
\begin{eqnarray}
\hat{\mathcal{I}}(t)=\Omega_{I}|\rho_0\rangle\rangle\langle\langle \text I|,\label{di1}
\end{eqnarray}
where $\Omega_I$ is an arbitrary nonzero constant and $\langle\langle  \text I|$ is the left vector corresponding to a $3\times 3$ identity matrix.

Assume that the total Liouvillian reads
\begin{eqnarray}
\hat{ \mathcal L}=\hat{ \mathcal L_c}+\Gamma_d\hat{ \mathcal D}[L_d].\label{cliou3}
\end{eqnarray}
The control Liouvillian has the same form as Eq.(\ref{me3}),
\begin{eqnarray}
\hat{ \mathcal L_c}\rho(t)&=&-i[{H}_c(t),{\rho}]\label{lc}\\
&&+\sum_{\alpha=\pm1}\Gamma_\alpha((N_\alpha^i+1)\hat {\mathcal D}[L_\alpha](\rho)
+N_\alpha^i\hat {\mathcal D}[L_\alpha^\dagger](\rho)),\nonumber
\end{eqnarray}
where the Hamiltonian is
\begin{eqnarray}
H_c(t)=\frac{\hbar}{2}
\left(
\begin{array}{ccc}
    0& \Omega_p^i(t)&i \,\Omega_c^i(t)\\
    \Omega_p^i(t)& 0& \Omega_s^i(t)\\
   - i\, \Omega_c^i(t)& \Omega_s^i(t)& 0\\
\end{array}
\right).
\end{eqnarray}
Here we also assume that the mean excitation numbers $N_{\pm 1}^i$ are  tunable independently, which can be achieved by properly engineering the temperature
of the environment \cite{Shabani2016} or shifting the energy difference between $\ket{A_2}$ and $\ket{m_s=\pm 1}$. Substituting Eqs. (\ref{di1}) and (\ref{lc}) into
Eq.(\ref{di}), we can determine all the control parameters in the control Liouvillian $\hat{ \mathcal L_c}$.  The analytical expression of these control parameters are
presented in Appendix. \ref{AD}. Alternatively, these control parameters can be obtained numerically, especially for more complex situation such as $\Gamma_{-1}
\neq\Gamma_{+1}$. Taking Eq.(\ref{ap3}) and Eqs.(\ref{r38m}) into the analytical expressions, we obtain the control parameters for the adiabatic trajectory  with a
time-independent $\Omega$,
\begin{eqnarray*}
&&\Omega_{s}^i=\Omega\,\sin(\theta(t)),\,\Omega_{p}^i=\Omega\,\cos(\theta(t)),\,\Omega_{c}^i=\partial_t \theta(t),\nonumber\\
&&N_{-1}^i=N,\,N_{+1}^i=N.\label{cpa3}
\end{eqnarray*}
These control parameters are  exactly the parameters obtained in the transitionless quantum driving scheme of  closed systems for the STIRAP\cite{Chen2010}.

Despite the MIE scheme and the transitionless driving scheme of closed systems provide similar control protocols for an adiabatic trajectory, their essences  are
quite different. The two schemes give similar control protocols only if $\Omega=\sqrt{\Omega_s^2(t)+\Omega_p^2(t)}$ is time-independent. This can be
illustrated by the spectrum decomposition of the instantaneous steady state of $\hat{ \mathcal L_0}$. The eigenvalues of $\rho_0$ are
 \begin{eqnarray}
p_1&=&\frac{1}{z}\left(\left(2\mathrm{N}+1\right)\left(3\mathrm{N}+2\right)\mathrm{N}\mathrm{\Gamma}^{2}+2\mathrm{N}\mathrm{\Omega}^{2}\right.\nonumber\\
&&\left.-\mathrm{N}\mathrm{\Gamma}\sqrt{\left(3\mathrm{N}+2\right)^{2}\mathrm{\Gamma}^{2}+4\mathrm{\Omega}^{2}}\right),\nonumber\\
p_2&=&\frac{1}{z}\left(\left(2\mathrm{N}+1\right)\left(3\mathrm{N}+2\right)\mathrm{N}\mathrm{\Gamma}^{2}+2\mathrm{N}\mathrm{\Omega}^{2}\right.\nonumber\\
&&\left.+\mathrm{N}\mathrm{\Gamma}\sqrt{\left(3\mathrm{N}+2\right)^{2}\mathrm{\Gamma}^{2}+4\mathrm{\Omega}^{2}}\right), \nonumber\\
p_3&=&\frac{\left(\mathrm{N}+1\right)}{z}\left(\left(3\mathrm{N}+2\right)\mathrm{N}\mathrm{\Gamma}^{2}+\mathrm{\Omega}^{2}\right).\nonumber
 \end{eqnarray}
with $z={\left(3\, \mathrm{N} + 1\right)\, {\mathrm{\Omega}}^2 + \mathrm{N}\, {\mathrm{\Gamma}}^2\, {\left(3\, \mathrm{N} + 2\right)}^2}$,
which denote the populations on corresponding eigenstates. If $\Omega$ is constant, the population on every eigenstate is invariant, so that a
unitary evolution is enough to transfer the quantum state into the target steady state.  But if $\Omega$ is time-dependent, the incoherent controls
are required for the STIRAP process along the adiabatic trajectory because of time-varying  purity of the quantum state.

\begin{figure}[htbp]
\centerline{\includegraphics[width=1.1\columnwidth]{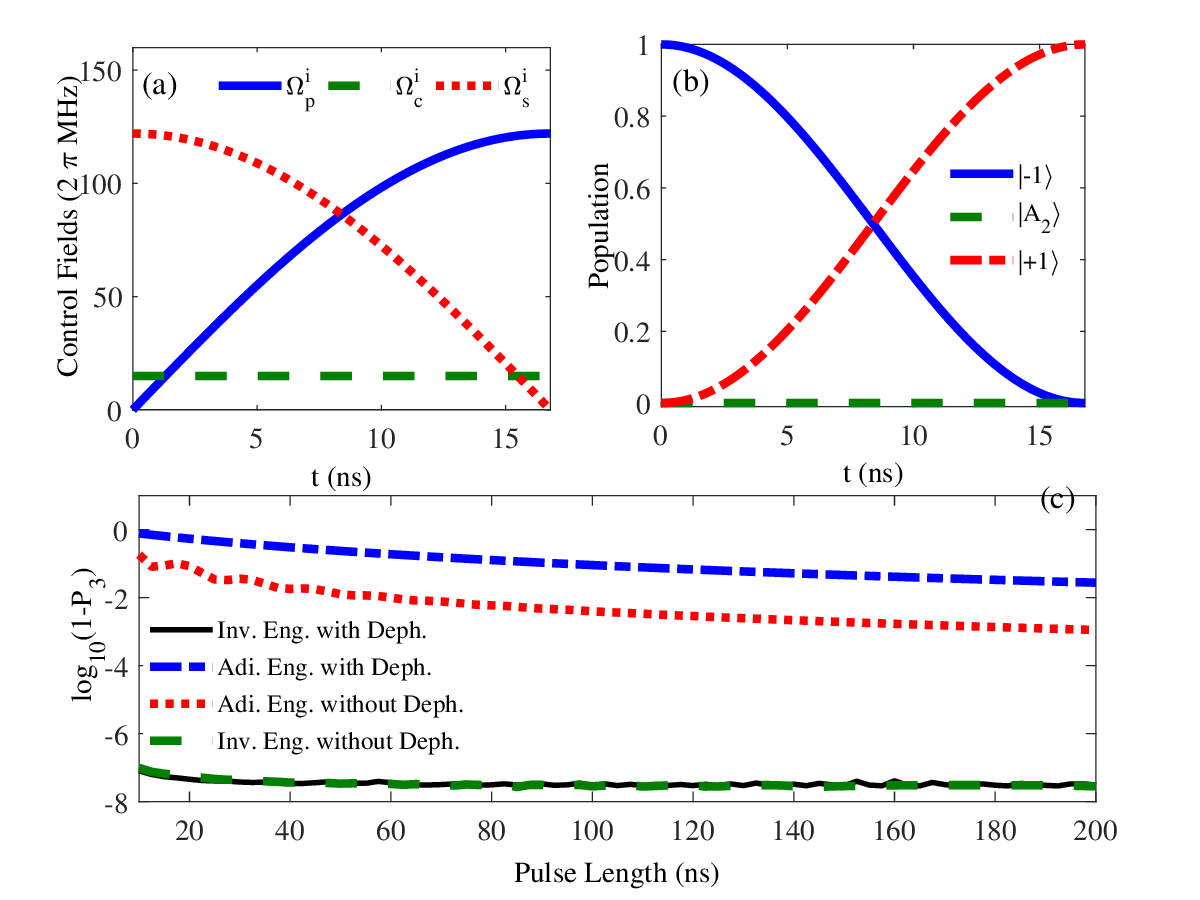}}
\caption{(a) The  control field and (b) the population on $\ket{m_s=-1}$ (blue solid line), $\ket{A_2}$ (green dash line),
and $\ket{+1}$ (red dot-dash line)  as a function of time for  pulses length $\tau=16.8$ ns, (c) the final population on states except  $\ket{m_s=+1}$
as a function of the pulse length $\tau$ for $\Omega=2\pi\times 122$ MHz, $\Gamma_{-1}=2\pi\times
4.3$ MHz, $\Gamma_{+1}=2\pi\times8.4$ MHz, $\Gamma_d=2\pi\times8.8$ MHz, and $N=1.9\times10^{-33}$.}\label{pp3}
\end{figure}

Figures \ref{pp3} (a) and (b) show the control fields  and the corresponding population for the open $\Lambda$ system at room temperature (300K). The corresponding
mean excitation number is $N=1.9\times10^{-33}$.  The boundary conditions of $\theta(t)$ are set to be $\theta(0)=0$ and $\theta(\tau)=\pi/2$. We choose the simplest
STIRAP pulses, i.e., $\theta(t)=\pi t/(2 \tau),$ where $\tau$ is the pulse length. The numerical results of the control fields show that, for nonadiabatic state transfer along  the
adiabatic trajectory, the control protocol is completely the same as the protocol given by the transitionless quantum driving method \cite{Chen2010}. This illustrates that the
transitionless driving method \cite{Vepsalainen2019,Zhang2013} will present a better transfer efficiency than any other scheme else\cite{Zhou2017,Du2016}, if $\Omega$
is set to be constant. As shown in  FIG.\ref{pp3} (b), the quantum state is transferred into the target state with a perfect final population. In FIG. \ref{pp3} (c), we plot the final
population not on $\ket{m_s=+1}$ (denoted by $\log_{10}(1-P_3)$ with $P_3=\bra{m_s=+1}\rho(\tau)\ket{m_s=+1}$) as a function of the pulse length $\tau$ for both the
adiabatic scheme (the red dots and blue dot-dash lines) and the MIE scheme (green dash and black solid lines). Comparing to the adiabatic scheme, our scheme can transfer
the population into $\ket{m_s=+1}$ with  transfer efficiency very close to 1(at least $10^{-6}$, see the figure). Since $\ket{A_2}$ is unoccupied, our protocol is immune to the
orbital dephasing  encountered in the earlier proposals.

In practical applications, the initial-to-final state coupling $\Omega_c^i$ can be implemented in some but not in all systems, e.g., because of selection rules due to
symmetry of the states or the necessary phase of the term. In nitrogen-vacancy electronic spins, this additional coupling was implemented mechanically via a strain
field in the recent experiment \cite{Barfuss2015,Kolbl2019}. The strain field to drive NV spins is based on the sensitive response of the NV spin states to strain in the
diamond host lattice. For uniaxial strain applied transverse to the NV axis, the transverse strain field couples the electronic spin states $\ket{m_s=-1}$ and
$\ket{m_s=+1}$. To realize such strain field for efficient coherent driving, a mechanical resonator is required in the form of a singly clamped, single-crystalline diamond
cantilever, in which the NV center is directly embedded \cite{Barfuss2015}. The cantilever is actuated at its mechanical resonance frequency $\omega_m=2\pi\times6.84$
MHz. Thus, for resonance driving, the external magnetic field satisfies $B_{NV}=2.42$  G. And the corresponding Rabi frequency is charactered by $\Omega_c=\gamma_
T x_{\text{c}}/x_{\text{zpf}}$, where $\gamma_T$ is the transverse single-phonon strain-coupling strength, $ x_{\text{c}}$ and $x_{\text{zpf}}$ are the cantilever
zero-point fluctuation and peak amplitude, respectively (with $\gamma_T\sim2\pi\times0.08$ MHz and $x_{\text{zpf}}\sim7.7\times10^{-15}$ m).
Therefore, we can tune the  cantilever’s  amplitude $ x_{\text{c}}$ to adjust the Rabi frequency of the strain  field.

\subsection{The Trajectory without Initial-to-Final State Couplings} \label{secIIIB}

In experiment, a initial-to-final state coupling induced by the strain field  $\Omega_c^i(t)$  can be realized  in artificial structure but not in real atoms via dipole-dipole coupling
due to the selection rule. The pure-state inverse engineering scheme solves this by providing alternative shortcuts that do not couple directly levels $\ket{m_s=-1}$ and
$\ket{m_s=+1}$ \cite{Chen2012}. In the following, we show that the MIE scheme has the same  quality via selecting a proper trajectory of the quantum state. And this particular
trajectory can be obtained only by solving differential equations about the components of the Bloch vector, but not to design a complex transformation of the entire trajectory as
done in the STAs schemes of closed quantum systems. More importantly, our protocol is robust to the decay and dephasing noise at the same time.

We consider the case where the Stocks pulse and the pumping pulse resonantly excite both $\ket{m_s=0}$ and $\ket{m_s=+1}$ to the excited-state spin singlet
$\ket{^1 E}$, as shown in FIG. \ref{il3} (c). The singlet-ground ($\ket{m_s=0}\rightarrow\ket{^1E}$) splitting is about 89 THz. The decay from the singlet state to
the ground states is  spin-nonpreservingnonradiative decay. The corresponding decay rates depend on the static magnetic field $\mathbf{B}_{NV}$ with an angle
$\eta$ with respect to the NV defect axis. It has been shown that the decay rates are approximatively equal for $\eta=\pi/10$, and have been measured in the lab with
the results $\Gamma_0\approx\Gamma_{+1}=2\pi\times1.5$ MHz \cite{Tetienne2012}. Moreover, due to spin-nonpreserving decay, the effective rate of the population
excitation from $\ket{m_s=0}$ to $\ket{^1 E}$ is much lower than the decay rate from $\ket{^1 E}$ to $\ket{m_s=0}$, while the effective rate from the $\ket{m_s=+1}$ to
$\ket{^1E}$ is almost equal to its corresponding decay rate. This results that the  $\ket{m_s=+1}-\ket{^1E}$ and  $\ket{m_s=0}-\ket{^1E}$ subsystems can be seen as
coupling to two different thermal reservoirs \cite{Klatzow2019}. It has also been  shown in Ref. \cite{Klatzow2019}  that tuning  the strength of $\mathbf{B}_{NV}$ can
effectively engineer temperatures of the  thermal reservoirs.

In order to cancel the initial-to-final state coupling, the control field $\Omega_c^i(t)$ must be zero at any point in time. A direct manner is to modify the trajectory of
the quantum state. Here, we consider the case where the decay rates are equal. As illustrated in the analytical expressions of the control parameters (see Appendix. \ref{AD}),
$\Omega_c^i(t)$ can be written as a function of the Bloch vector and its time derivative $\{r_i,\,\partial_t r_i\}_{i=1}^8$.  What we may do is   to force $r_4(t)$ to vary in
a proper way such that $\Omega_c^i(t)=0$. To be specific, we consider the requirement $\Omega_c^i(t)=0$  as a restriction to solve the differential equation of $r_4(t)$,
keeping other components of the Bloch vector unchanged. This would finally leads to  a proper trajectory that cancels  the initial-to-final state coupling.

\begin{figure}[htbp]
\centerline{\includegraphics[width=1.1\columnwidth]{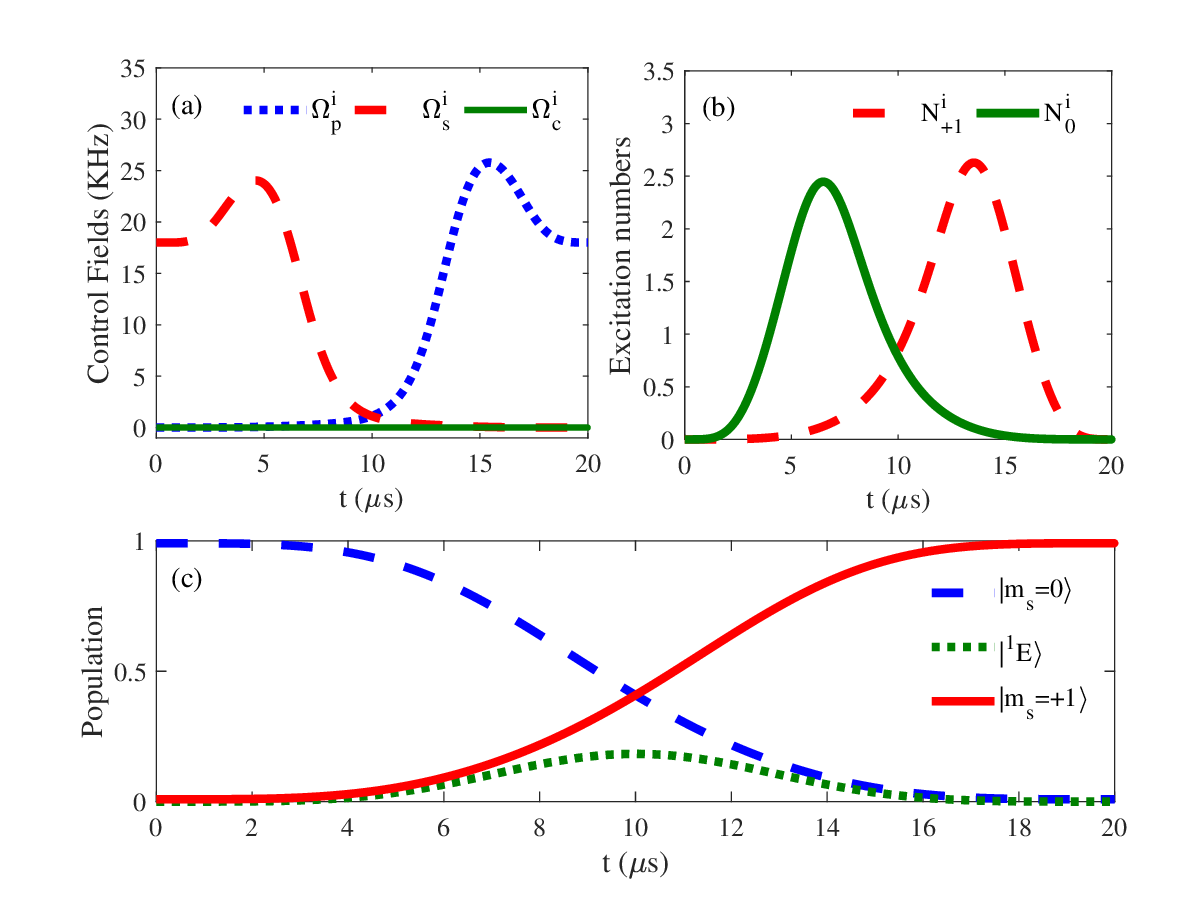}}
\caption{(a) The control fields,  (b) the mean excitation numbers, and (c) the population on states $\ket{m_s=0}$ (blue lines), $\ket{^1 E}$ (green lines),
and $\ket{m_s=+1}$ (red lines)  as a function of time for $\Omega=18.1$ KHz,  $\Gamma=2\pi\times1.5$ MHz,
$\Gamma_d=2\pi\times8.8$ MHz, and $N=6.55\times10^{-7}$. The pulses length is $\tau=20\,\mu\text{s}$.}\label{Lam3}
\end{figure}

By choosing a trajectory according to the instantaneous steady state of $\hat{ \mathcal L_0}$ (see Appendix. \ref{AE}), we plot the control parameters and the populations as
a function of time in FIG. \ref{Lam3}. We notice that the additional coupling $\Omega_c^i(t)$ is eliminated  in the engineering process (the green solid line in FIG. \ref{Lam3} (a)),
and the population is transferred from $\ket{m_s=0}$ into the target state $\ket{m_s=+1}$ with high transfer efficiency as  shown in  FIG. \ref{Lam3} (c). In addition,
 the excitation numbers of the  reservoirs need to be engineered accordingly (see  FIG. \ref{Lam3} (b)).

\begin{figure}[htbp]
\centerline{\includegraphics[width=1.1\columnwidth]{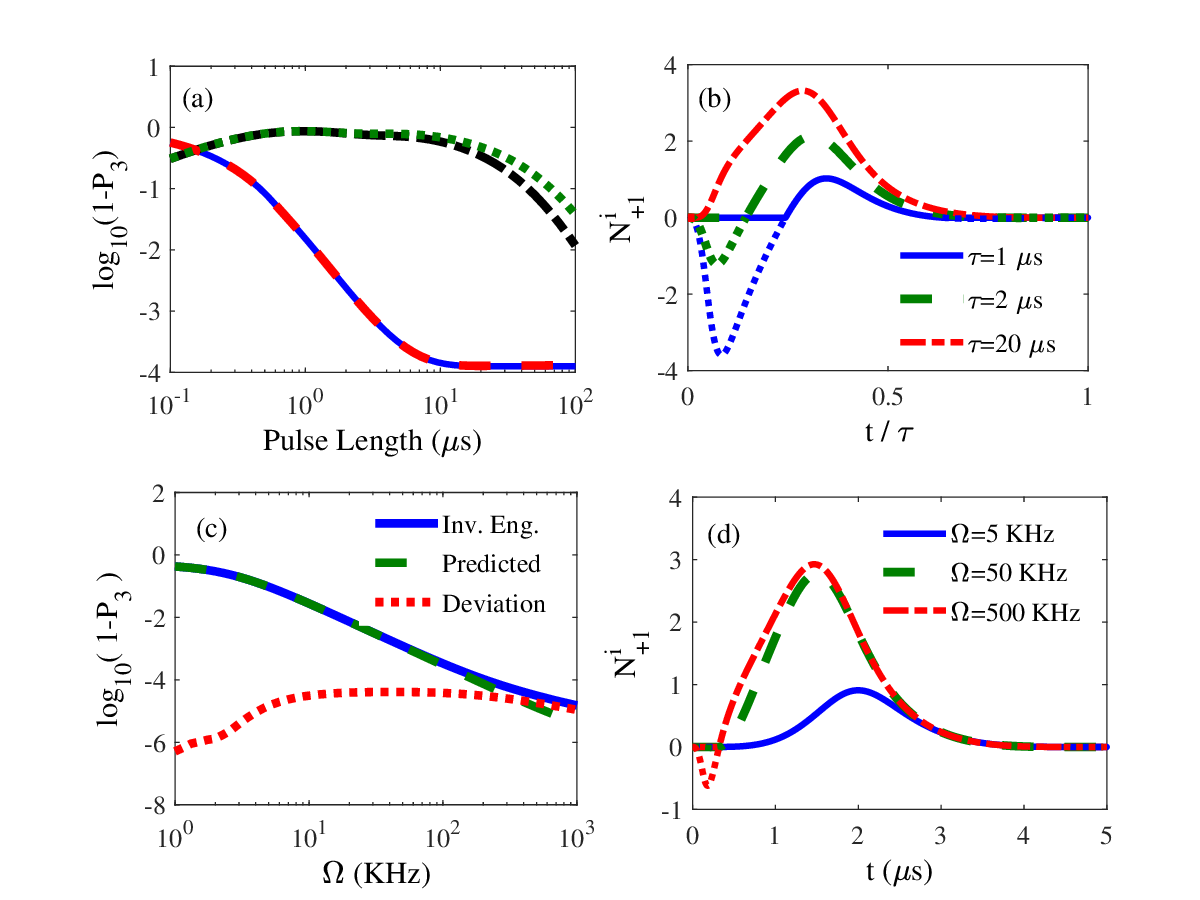}}
\caption{ The final population not on $\ket{m_s=+1}$ (red lines) as a function of  (a) the pulse length $\tau$ for $\Omega=154$ KHz and (c)
the control fields $\Omega$ for $\tau=5 \,\mu$s. The main excitation number $N_{+1}^i$ as a function of time with (b) different pulse
length for $\Omega=154$ KHz and (d) different control fields for $\tau=5 \,\mu$s. The other parameters in the master equation are chosen as
 $\Gamma=2\pi\times1.5$ MHz, $\Gamma_d=2\pi\times8.8$ MHz, and $N=6.55\times10^{-7}$.}\label{PO3}
\end{figure}

Although FIG. \ref{Lam3} presents  positive main excitation numbers, these main excitation
numbers may still negative at some points of the time. In experiment, we may restrict  the main excitation numbers to be within the regime $N_{0}^i,\,
N_{+1}^i\geq6.55\times10^{-7}$ which corresponds to  $T_{\text{cutoff}}\approx 300$ K. In FIG. \ref{PO3}(a), we plot the final
population which is not on $\ket{m_s=+1}$ as a function of the pulse length $\tau$ for the control field $\Omega=154$ KHz. Both dynamical processes
with (the red dash line)  and without (the blue solid line) the dephasing noise are considered in FIG.\ref{PO3}.
As we see, the MIE scheme  fails to transfer the population into $\ket{m_s=+1}$ for short pulse length
due to the cut-off on the main excited number $N_{+1}^i$. But it performs good
 by prolonging pulse length, and the final population on $\ket{m_s=+1}$  approach asymptotically  to a predicted value give by
 the instantaneous steady state of $\hat{ \mathcal L_0}(\tau)$.  On the other hand, since the coherence between $\ket{^1E}$ and
 $\ket{m_s=0}$ ($\ket{m_s=+1}$) in our designed trajectory is negligible, the orbital dephasing noise of $\ket{^1 E}$ will not
affect the state transfer process evidently. The numerical result confirms our analysis  as illustrated by the blue solid line and the red dash line in
FIG. \ref{PO3} (a). We also present the results given by the superadiabatic scheme\cite{Barfuss2015} with  (the green dots line) and without  (the black dash-dot line)
the dephasing noise. As a pure-state STAs scheme, the superadiabatic scheme needs to keep states as pure states. When $\ket{^1 E}$ is populated,
the strong coherence between the excited and ground states are required, so that pure-state STAs schemes are sensitive to the dephasing noise.
Therefore, the MIE scheme is more robust to  the orbital dephasing noise than the STAs schemes of closed quantum systems.

We also plot the  population on all states except  $\ket{m_s=+1}$(blue solid line) as a function of the control field $\Omega$ with a pulse length
$\tau=5\, \mu\text{s}$  in FIG. \ref{PO3} (c). As expected, the final population on $\ket{m_s=+1}$ increases  with the control field.  The  population
not on $\ket{m_s=+1}$ given by the steady state $\rho_0(\tau)$ is also plotted, see the green dash line. We find that the final population for the inverse engineering
 can not reach the predicted value due to the main excitation number cutoff, see  FIG. \ref{PO3} (d).
As illustrated by the red dot line in FIG. \ref{PO3} (c), the worst deviation of the final population is no more than $10^{-4}$. Therefore, although the
 final population deviates from the predicted value, we can still obtain a satisfactory transfer efficiency in the regime of strong control field.

\section{Comparison with the Pure-state Inverse Engineering Scheme} \label{comparison}

 \subsection{Pure-state Engineering}

In this section, we compare the MIE scheme to the pure-state inverse engineering (PIE) scheme. Firstly, we show that the STAs scheme of open quantum systems
includes  the PIE scheme.  To inversely engineering a pure-state, the dynamical invariants $I(t)$ of closed systems are needed\cite{Chen2011}, which satisfies
\begin{eqnarray*}
i\,\partial_t I(t)=[H(t),I(t)],
\end{eqnarray*}
 with the Hamiltonian $H(t)$. The general solution of closed systems can be written as
\begin{eqnarray}
\ket{\Psi(t)}=\sum_n c_n \exp(i\alpha_n(t))\ket{\psi_n(t)}, \label{gs}
\end{eqnarray}
where $c_n$ are time-independent amplitudes,  $\ket{\psi_n(t)}$ are orthonormal eigenvectors of the invariant $I(t)$, and $\alpha_n(t)$ are the Lewis-Riesenfeld phases.
 According to the general solution Eq.(\ref{gs}), we can write down the time-dependent unitary evolution operator $U$ as
\begin{eqnarray*}
U(t)=\sum_n \exp(i\alpha_n(t))\ket{\psi_n(t)}\bra{\psi_n(0)},
\end{eqnarray*}
which obeys
\begin{eqnarray*}
i\partial_t U(t)=H(t)U(t).
\end{eqnarray*}
Thus, we obtain the formal expression of the Hamiltonian as
\begin{eqnarray}
H(t)=\sum_n\left(i\,\ket{\partial_t\psi_n(t)}\bra{\psi_n(t)}-\partial_t\alpha_n(t)\ket{\psi_n(t)}\bra{\psi_n(t)}\right).\nonumber\\\label{ht}
\end{eqnarray}
If we impose $[H(0),I(0)]=[H(\tau),I(\tau)]=0$, the eigenstates coincide and then a state transfer from a initial eigenstate
of the Hamiltonian to the final one is guaranteed.

For applying the STAs scheme of open quantum systems to a pure-state engineering task, we can formulate the dynamical invariant
superoperator $\hat{\mathcal I}$ by the eigenstates of closed systems' dynamical invariant. The time-dependent eigenstates of
$\hat{\mathcal I}$ are chosen as
\begin{eqnarray}
|\Psi_{mn}\rangle\rangle=\ket{\psi_m}\otimes\ket{\psi_n^*},\label{pmn}
\end{eqnarray}
which corresponds to the time-independent eigenvalues $\lambda_{mn}$. Substituting  $\{|\Psi_{mn}\rangle\rangle\}$ into Eq.(\ref{contliou}),
the control Liouvillian superoperator can be obtained with arbitrary phases $\eta_{mn}(t)$. Thus, $\hat{\mathcal L}_c(t)$ transfers the
quantum state from $\ket{\psi_n(0)}$ to $\ket{\psi_n(\tau)}$ along with $\ket{\Psi_{nn}(t)}\rangle$. According to Eq.(\ref{contliou}), the control
Liouvillian can be written as
\begin{eqnarray*}
\hat{\mathcal L}_c(t)=\sum_{mn}\left(|\partial_t\Psi_{mn}\rangle\rangle\langle\langle\Psi_{mn}|+\partial_t\eta_{mn}|\Psi_{mn}\rangle
\rangle\langle\langle\Psi_{mn}|\right).
\end{eqnarray*}
By considering Eq.(\ref{pmn}), it yields
\begin{eqnarray*}
\hat{\mathcal L}_c(t)&=&\sum_{mn}\left(|\partial_t\psi_{m}\rangle\langle\psi_{m}|\otimes(|\psi_{n}\rangle\langle\psi_{n}|)^*\right.\nonumber\\
&&+|\psi_{m}\rangle\langle\psi_{m}|\otimes(\partial_t|\psi_{n}\rangle\langle\psi_{n}|)^*\nonumber\\
&&\left.+\partial_t\eta_{mn}|\Psi_{mn}\rangle\rangle\langle\langle|\Psi_{mn}|\right).
\end{eqnarray*}
Since $\{\ket{\psi_m}\}$ is a complete set of the Hilbert space, we have
\begin{eqnarray*}
&\hat{\mathcal L}_c(t)&=-i\left(\right.i\,\sum_{mn}\partial_t\eta_{mn}|\Psi_{mn}\rangle\rangle\langle\langle|\Psi_{mn}|\\
&&\left.+(\sum_{m}i\,|\partial_t\psi_{m}\rangle\langle\psi_{m}|)\otimes\text I
-\text I\otimes(\sum_{n}i\,\partial_t|\psi_{n}\rangle\langle\psi_{n}|)^*\right).\nonumber
\end{eqnarray*}
where $\text I$ is an  identity matrix. If we choose the phase satisfies $\eta_{mn}=i\,(\alpha_m-\alpha_n)$,
it can be obtained that
\begin{eqnarray*}
\hat{\mathcal L}_c(t)&&=-i\left((i\,\sum_{m}|\partial_t\psi_{m}\rangle\langle\psi_{m}|-\partial_t \alpha_m|\partial_t\psi_{m}
\rangle\langle\psi_{m}|)\otimes \text I\right.\nonumber\\
&&\left.-\text I\otimes(i\,\sum_{n}\partial_t|\psi_{n}\rangle\langle\psi_{n}|-\partial_t \alpha_n|\partial_t\psi_{n}\rangle\langle\psi_{n}|)^*\right).
\end{eqnarray*}
Considering the map from the superoperator to operators: $A\otimes B^*|\rho\rangle\rangle\mapsto A\rho B^\dagger$ for arbitrary operators
$A$ and $B$, we immediately have
\begin{eqnarray}
\hat{\mathcal L}_c(t)|\rho(t)\rangle\rangle=-i[H(t),\rho(t)],
\end{eqnarray}
where $H(t)$ is just the Hamiltonian involved in the PIE scheme as shown in  Eq.(\ref{ht}). Therefore, the STAs scheme of open quantum systems includes
the PIE scheme.

In the following, we take the STIRAP as an example to shown that the STAs scheme gives same
control protocol as the PIE scheme. Following the step of Ref.\cite{Chen2012}, the eigenstates
of dynamical invariant are given by
\begin{eqnarray*}
&&\ket{\psi_0}=\left(
               \begin{array}{c}
                 \cos\gamma\,\cos\beta \\
                 -i\sin\gamma \\
                 -\cos\gamma\,\sin\beta \\
               \end{array}
             \right),\,\nonumber\\
&&\ket{\psi_\pm}=\frac{\sqrt{2}}{2}\left(
               \begin{array}{c}
                 \sin\gamma\,\cos\beta\pm i \sin\beta \\
                 i\cos\gamma \\
                 -\sin\gamma\,\sin\beta \pm i\cos\beta\\
               \end{array}
             \right).
\end{eqnarray*}
Thus, we can formulate the eigenstate of the dynamical invariant superoperator according to Eq.(\ref{pmn}). Substituting Eq.(\ref{pmn}) into
Eq.(\ref{contliou}), we obtain the control Liouviilian $\hat{\mathcal L}_c$ as a function of
$\beta$, $\gamma$, and phases $\eta_{mn}$. Comparing $\eta_{mn}$ with the Lewis-Riesenfeld phases $\alpha_n$, we immediately find
that $\eta_{mn}=i(\alpha_m-\alpha_n)$. Thus, we have $\eta_{nn}=0$,  $\eta_{+-}=-\eta_{-+}\equiv\eta_1$, $\eta_{+0}=-\eta_{0+}\equiv\eta_2$,
and $\eta_{0-}=-\eta_{-0}\equiv\eta_3$.  The control Liouvillian can be expanded by SU(3) generators ($T_i$ - the regular Gellmann matrixes Eq.(\ref{Tx})), i.e.,
\begin{eqnarray*}
\hat{\mathcal L}_c(t)=\sum_{i,j=1}^9 c_{ij} T_i\otimes T_j^*.
\end{eqnarray*}
where $T_9$ is a $3\times 3$ identity matrix. $c_{ij}$ is time-dependent expanding coefficients, which can be determined by
\begin{eqnarray*}
 c_{ij}=\text{Tr}(\hat{\mathcal L}(t) T_i\otimes T_j^*).
\end{eqnarray*}
Moreover, if  the phase $\eta_k\,(k=1,2,3)$ have following relation: $\eta_2=\eta_1/2$ and $\eta_3=-\eta_1/2$, it yields the coefficient matrix
\begin{widetext}
\begin{eqnarray}
&&c=\frac{i}{2}\times\nonumber\\&&\left(\begin{array}{ccccccccc} 0 & 0 & 0 & 0 & 0 & 0 & 0 & 0 &  - 2\partial_t \gamma\cos\beta - \eta_1 \cos\gamma\sin\beta\\ 0 & 0 & 0 & 0 & 0 & 0 & 0 & 0 & 0\\ 0 & 0 & 0 & 0 & 0 & 0 & 0 & 0 & 0\\ 0 & 0 & 0 & 0 & 0 & 0 & 0 & 0 & 0\\ 0 & 0 & 0 & 0 & 0 & 0 & 0 & 0 & 2\partial_t \beta  - \eta_1\sin\gamma\\ 0 & 0 & 0 & 0 & 0 & 0 & 0 & 0 & 2 \partial_t \gamma\sin\beta - \eta_1\cos\beta \cos\gamma\\ 0 & 0 & 0 & 0 & 0 & 0 & 0 & 0 & 0\\ 0 & 0 & 0 & 0 & 0 & 0 & 0 & 0 & 0\\ 2\partial_t \gamma\cos\beta + \eta_1\cos\gamma\sin\beta & 0 & 0 & 0 &  - 2\partial_t \beta + \eta_1\sin\gamma &  -2\partial_t \gamma \sin\beta+ \eta_1 \cos\beta \cos\gamma & 0 & 0 & 0 \end{array}\right).\nonumber
\end{eqnarray}
\end{widetext}
In other words, the control Liouvillian reads
\begin{eqnarray*}
\hat{\mathcal L}_c(t)=&&-i\left((2\partial_t \gamma\cos\beta+\eta_1 \cos\gamma\sin\beta)(T_1\otimes T_9^*-T_9\otimes T_1^*)\right.\nonumber\\
&&+(2\partial_t \beta  - \eta_1\sin\gamma)(T_5\otimes T_9^*-T_9\otimes T_5^*)\nonumber\\
&&\left.+(\eta_1\cos\beta \cos\gamma-2 \partial_t \gamma\sin\beta)(T_6\otimes T_9^*-T_9\otimes T_6^*)\right).
\end{eqnarray*}
Thus, the control Liouvillian can be transformed as
\begin{eqnarray}
\hat{\mathcal L}_c(t)|\rho(t)\rangle\rangle\mapsto -i[H(t),\rho(t)]
\end{eqnarray}
with $H(t)=\Omega_p/2\, T_1+\Omega_c/2\, T_5+\Omega_s/2\, T_6$, in which the control fields are
\begin{eqnarray}
\Omega_p&=&\eta_1 \cos\gamma\sin\beta+2\partial_t \gamma\cos\beta,\nonumber\\
\Omega_c&=&2\partial_t \beta  - \eta_1\sin\gamma,\nonumber\\
\Omega_s&=&\eta_1\cos\beta \cos\gamma-2 \partial_t \gamma\sin\beta.
\end{eqnarray}
In order to cannel  initial-to-final state couplings, the phase $\eta_1$ has to selected as $\eta_1=2\partial_t \beta/\sin\gamma$. Taking $\eta_1$ into
above equations, we immediately obtain the same control protocol given in Ref.\cite{Chen2012}, i.e.,
\begin{eqnarray}
\Omega_p&=&2\partial_t \beta \cot\gamma\sin\beta+2\partial_t \gamma\cos\beta,\nonumber\\
\Omega_s&=&2\partial_t \beta\cot \gamma\cos\beta-2 \partial_t \gamma\sin\beta,\nonumber\\
\Omega_c&=&0.\label{pcp}
\end{eqnarray}
Therefore, the STAs scheme of open quantum systems is equivalent to the PIE scheme if the control task is to transfer pure states with a pure-state trajectory.

\subsection{Pure-state Engineering with Mixed-state Trajectories}

\begin{figure}[htbp]
\centerline{\includegraphics[width=1.1\columnwidth]{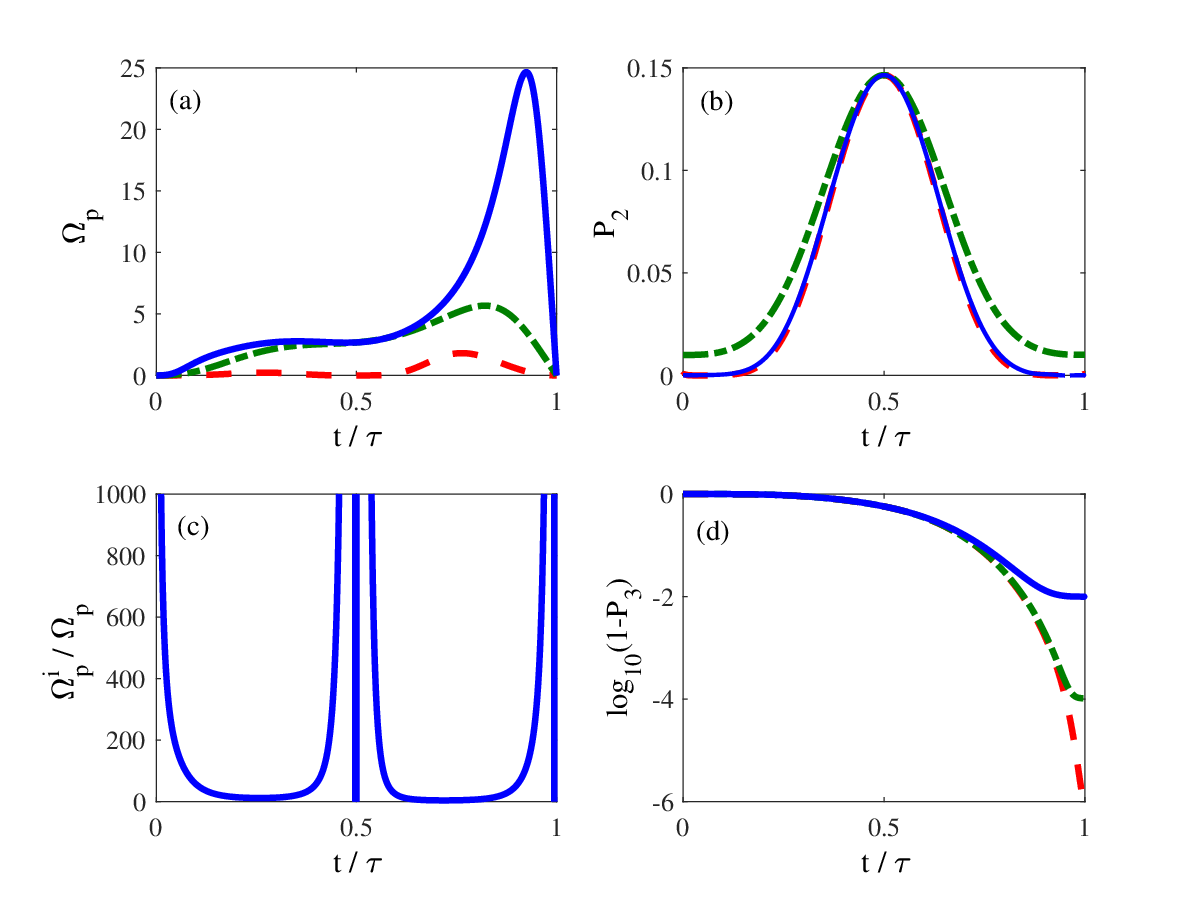}}
\caption{ (a) The control field $\Omega_p$, (b) the population on $\ket{^1E}$, (c) the ratio of the control field strengths between the mixed-state
and the pure-state inverse engineering schemes, and (d) the population out of $\ket{+1}$ vs the  dimensionless time $t/\tau$ for
the mixed-state trajectory (red dash lines) and the pure state trajectory with $\gamma_0=0.1$ (green dot-dash lines) and $\gamma_0=0.01$
(blue solid lines). The decay rate is chosen as $\Gamma=10/\tau$. }\label{mpc}
\end{figure}

The MIE scheme provides  more feasible control protocols than the PIE scheme, because the trajectory does not has to be a pure-state
trajectory. As illustrated by the STRIAP of open quantum systems, the robustness to the dephasing noise attributes to the mixed-state
trajectory with weak coherence between the energy levels. In this subsection, we show that the MIE scheme overcomes the difficulties
meeting in the PIE scheme. As shown in Eq.(\ref{pcp}), $\ket{^1E}$ needs to be populated for avoiding infinitely large $\Omega_{p,s}$.
The strengths of control fields satisfy $\Omega_{p,s}\propto 1/\sqrt{P_2}$ where $P_2=\sin^2\gamma$ is the population on  $\ket{^1E}$.
For the control protocol given by the MIE scheme,  reasonable and feasible   $\Omega_{p,s}$ are needed instead of infinitely large  control fields,
which is illustrated in FIG. \ref{mpc} (a).

For the control protocol given by the PIE scheme, the parameters in Eq.(\ref{pcp}) is chosen as: $\gamma=(\gamma_m-\gamma_0)
\sin^3(\pi t/\tau)+\gamma_0$ with constants $\gamma_m$ and $\gamma_0$; $\beta=\pi/2\sin(\pi t/(2\tau))$. For the mixed-state protocol,
we set a analogous trajectory with $r_3=-\sqrt{3}(\sin^2\gamma- \cos^2\beta\cos^2\gamma)/2$, $r_8=(3\sin^2\gamma + 3\cos^2\beta\cos^2
\gamma - 2)/2$. Since the mixed-state protocol does not need the population on $\ket{^1 E}$, we set $\gamma_0=0$ in the mixed-state trajectory.
The other components of the bloch vector are $r_{2}=-\sin^2\xi\cos\beta$ and $r_7=\sin^2\xi\sin\beta$ with $\xi=\xi_0\sin^2(2\pi t/\tau)$;
$r_4$ is used to cannel the initial-final state coupling, which is determined by the equation $\Omega^i_c(r_4)=0$.

In FIGs. \ref{mpc} (a) and (b), we plot the control field $\Omega_p$ and the population on $\ket{^1 E}$ as a function of the dimensionless time
$t/\tau$. The red dash lines are the results given by the mixed-state protocol, and the green dot-dash lines and the blue solid lines are the numerical
results given by the pure-state protocol with $\gamma_0=0.1$ and $\gamma_0=0.01$ respectively. As shown in FIG.\ref{mpc} (a), the mixed-state
protocol only requires finite strength of the control field  $\Omega_p^i$, even if  the population on $\ket{^1 E}$ is zero. But, for the PIE scheme, the strength of
$\Omega_p$ tends to infinity when the population on $\ket{^1 E}$ goes to zero at $t/\tau=1$. The finite control field strength is the contribution from
designable $r_2$ and $r_7$. For $t/\tau=1$, it can be obtained that the final state satisfies $r_3=0$, $r_8=-1$ and $r_4=0$, and the time
derivative of these components of the general Bloch vector are zeros. Then, we can obtain the control field $\Omega_p^i$ as a function of
$r_2$ and $r_7$ from Eq.(\ref{cp3}), which reads
\begin{eqnarray}
\Omega_p^i=-\frac{2\sqrt{3}\,\partial_tr_7\,r_2 -\sqrt{3}\,\partial_tr_2\,r_7 + 2\,\sqrt{3}\,\Gamma\,r_2\,r_7}{2\,r_2\,(- r_2^2 + r_7^2 + 3)}.\label{limit1}
\end{eqnarray}
This equation illustrates that, even if $|^1E\rangle$ is unpopulated, we can still obtain a finite control field by selecting proper relation
between $r_2$ and $r_7$. Substituting the concrete parameterized $r_2$ and $r_7$ into Eq.(\ref{limit1}), we finally obtain $\Omega_p^i(\tau)=0$.
To illustrate this clearly, we further plot the ratio between the control fields for the mixed-state protocol $\Omega_p^i(t)$ and the pure-state protocol
$\Omega_p(t)$ with $\gamma_0=0.01$ in FIG. \ref{mpc} (c). When $t/\tau\rightarrow0$ and  $t/\tau\rightarrow1$ (the
minimal population on $\ket{^1 E}$ for the pure-state protocol), the ratio goes to infinity, which verifies that the mixed-state protocol does not
require a extremely strong control field in the control process. As shown in FIG.\ref{mpc} (d), the MIE scheme also presents a better transfer efficiency
than the pure-state inverse engineering scheme.

\section{Conclusion}\label{conclusion}

To achieve feasible control of open quantum systems with high-accuracy, high-controllability, and  high-speed, we propose a fast and robust
control scheme. After presenting the STAs  based on dynamical invariants of open quantum system, we apply STAs
of open quantum systems to accelerate the adiabatic steady process. As a result, with the same form  as the reference Liouvillian,
the control Liouvillian can drive the open quantum system from an initial steady state into a target steady  state  along a designed trajectory with
desired fidelity and pulse length. We highlight the high-controllability of the trajectory, which leads to robustness of control protocols to some
particular noises and eliminates untunable control manners.

Our scheme opens several promising avenues for further developments. Theoretically, it would be interesting to explore possible speed-limits
and trade-off relations for the open quantum systems \cite{Funo2019,Pietzonka2018}. Experimentally, due to the feasible of the MIE protocol,
the present protocol can be realized in various systems, such as cavity quantum electrodynamical systems\cite{Bason2012}, superconducting circuits
\cite{Vepsalainen2019}, nitrogen-vacancy centers \cite{Zhang2013} and spin-chains\cite{Zhou2020}.

 This work is supported by National Natural Science Foundation of China (NSFC) under Grants No.
12075050 and 11775048.

\appendix

\section{The General Solution of Eq.(\ref{deq1})} \label{AA}

{The dynamical invariants $\hat{\mathcal I}(t)$ are defined as superoperators which satisfy the dynamical equation
\begin{eqnarray}
\partial_t \hat{\mathcal I}(t)-[\hat{\mathcal L}_c(t),\hat{\mathcal I}(t)]=0,\label{a10}
\end{eqnarray}
where $\hat{\mathcal L}(t)$ is the control Liouvillian superoperator.
Generally speaking, the superoperator $\hat {\mathcal I}(t)$ is non-Hermitian. We can introduce
a right basis $\{|D_\alpha^{(i)}\rangle\rangle\}$ and left basis $\{\langle\langle E_\alpha^{(i)}|\}$
in Hilbert-Schmidt space based on the Jordan canonical form. The left and right
basis always satisfy
\begin{eqnarray}
\hat {\mathcal I}\,|D_\alpha^{(i)}\rangle\rangle=\lambda_\alpha|D_\alpha^{(i)}\rangle\rangle+|D_\alpha^{(i-1)}\rangle\rangle,\label{a11}\\
\langle\langle E_\alpha^{(i)}| \hat {\mathcal I}=\lambda_\alpha\langle\langle E_\alpha^{(i)}|+\langle\langle E_\alpha^{(i+1)}|,\nonumber
\end{eqnarray}
where
\begin{eqnarray}
|D_\alpha^{(-1)}\rangle\rangle\equiv0,\label{rmax}\\
\langle\langle E_\alpha^{(n_\alpha)}|\equiv0\label{lmax},
\end{eqnarray}
with $i=0,1,...,n_\alpha-1$ ($n_\alpha$ is the dimension of the block $\alpha$). Here we assume that all of eigenvalues are nondegenerate, i.e.,
$\lambda_\alpha\neq\lambda_\beta$ for $\forall\, \alpha\neq\beta$. Moreover, the left and right bases  satisfy the orthonormality condition
\begin{eqnarray*}
\langle\langle E_\alpha^{(i)}|D_\beta^{(j)}\rangle\rangle=\delta_{\alpha\beta}\delta_{ij}
\end{eqnarray*}
The right and left basis $|D_\alpha^{(0)}\rangle\rangle$ and $\langle\langle E_\alpha^{(n_\alpha-1)}|$ are the right and left eigenvector of $\hat {\mathcal I}(t)$
with the eigenvalue $\lambda_\alpha$. Taking the first  derivative of Eq.(\ref{a11}) with respect to time, we have
\begin{eqnarray}
\partial_t\,\hat {\mathcal I}\,|D_\alpha^{(i)}\rangle\rangle+&\hat {\mathcal I}&\,|\partial_t\,D_\alpha^{(i)}\rangle\rangle
=\partial_t\lambda_\alpha|D_\alpha^{(i)}\rangle\rangle\nonumber\\
&+&\lambda_\alpha|\partial_tD_\alpha^{(i)}\rangle\rangle+|\partial_t
D_\alpha^{(i-1)}\rangle\rangle.
\end{eqnarray}
Multiplying  above equation and  $\langle\langle E_\beta^{(j)}|$ yields
\begin{eqnarray}
\langle\langle &E_\beta^{(j)}&|\partial_t\,\hat {\mathcal I}\,|D_\alpha^{(i)}\rangle\rangle=\partial_t\lambda_\alpha \delta_{\alpha\beta}
\delta_{ij}\nonumber\\
&&+(\lambda_\alpha-\lambda_\beta)\langle\langle E_\beta^{(j)}| \partial_tD_\alpha^{(i)}\rangle\rangle+\langle\langle E_\beta^{(j)}
|\partial_tD_\alpha^{(i-1)}\rangle\rangle\nonumber\\
&&-\langle\langle E_\beta^{(j+1)}|\partial_tD_\alpha^{(i)}\rangle\rangle.
\end{eqnarray}
Substituting  Eq.(\ref{a10}) into the last equation, we obtain
\begin{eqnarray}
\partial_t\lambda_\alpha &\delta_{\alpha\beta}&\delta_{ij}=(\lambda_\alpha-\lambda_\beta)\langle\langle E_\beta^{(j)}| \hat O|D_\alpha^{(i)}\rangle\rangle\nonumber\\
&&+\langle\langle E_\beta^{(j)}|\hat O|D_\alpha^{(i-1)}\rangle\rangle-\langle\langle E_\beta^{(j+1)}|\hat O|D_\alpha^{(i)}\rangle\rangle,\label{lambdat}
\end{eqnarray}
where $\hat O\equiv\hat{\mathcal L}_c-\partial_t$.}

For $\alpha\neq\beta$, the above equation yields
\begin{eqnarray}
(\lambda_\alpha-&\lambda_\beta&)\langle\langle E_\beta^{(j)}| \hat O|D_\alpha^{(i)}\rangle\rangle
+\langle\langle E_\beta^{(j)}|\hat O|D_\alpha^{(i-1)}\rangle\rangle\nonumber\\
&&-\langle\langle E_\beta^{(j+1)}|\hat O|D_\alpha^{(i)}\rangle\rangle=0,\label{recurrence}
\end{eqnarray}
In the following, we illustrate that $\langle\langle E_\beta^{(j)}| \hat O|D_\alpha^{(i)}\rangle\rangle=0$ for $\forall\,i,\,j$. Firstly, we notice that
above equation is a recurrence equation about the indexes $i$ and $j$. Setting $i=0$ and $j=n_\beta-1$, we obtain
\begin{eqnarray}
\langle\langle E_\beta^{(n_\beta-1)}| \hat O|D_\alpha^{(0)}\rangle\rangle=0,\label{n0x}
\end{eqnarray}
 where Eqs. (\ref{rmax}) and (\ref{lmax}) are used. Then,  using the recurrence equation Eq.(\ref{recurrence}) again and considering $i=1$ and $j=n_\beta-1$, it results in
 \begin{eqnarray*}
(\lambda_\alpha-\lambda_\beta)\langle\langle E_\beta^{(n_\beta-1)}| \hat O|D_\alpha^{(1)}\rangle\rangle
+\langle\langle E_\beta^{(n_\beta-1)}|\hat O|D_\alpha^{(0)}\rangle\rangle=0,
\end{eqnarray*}
Placing Eq.(\ref{n0x}) into above equation, we have
\begin{eqnarray}
\langle\langle E_\beta^{(n_\beta-1)}| \hat O|D_\alpha^{(1)}\rangle\rangle=0,\label{n1}
\end{eqnarray}
Repeating this procedure and checking every index $i=1,...,n_\alpha-1$, we conclude that the follow relations are insured,
\begin{eqnarray}
\langle\langle E_\beta^{(n_\beta-1)}| \hat O|D_\alpha^{(i)}\rangle\rangle=0,\,\forall i=0,...,n_\alpha-1.\label{nxx}
\end{eqnarray}
Secondly, we check $\langle\langle E_\beta^{(j)}| \hat O|D_\alpha^{(0)}\rangle\rangle$ by the same procedure as before. Setting $j=n_\beta-2$, Eq.(\ref{recurrence}) is turning into
\begin{eqnarray*}
(\lambda_\alpha-\lambda_\beta)\langle\langle E_\beta^{(n_\beta-2)}| &\hat O&|D_\alpha^{(0)}\rangle\rangle
+\langle\langle E_\beta^{(n_\beta-2)}|\hat O|D_\alpha^{(-1)}\rangle\rangle\nonumber\\
&&-\langle\langle E_\beta^{(n_\beta-1)}|\hat O|D_\alpha^{(0)}\rangle\rangle=0.
\end{eqnarray*}
Due to the relation Eqs. (\ref{rmax}) and (\ref{n0x}), we immediately have
\begin{eqnarray*}
\langle\langle E_\beta^{(n_\beta-2)}| \hat O|D_\alpha^{(0)}\rangle\rangle=0.
\end{eqnarray*}
Thus, reducing the index $j$ step by step and check all terms by Eq.(\ref{recurrence}), it can be concluded that
\begin{eqnarray}
\langle\langle E_\beta^{(j)}| \hat O|D_\alpha^{(0)}\rangle\rangle=0.\label{nyy}
\end{eqnarray}
Equipping with Eqs. (\ref{nxx}) and (\ref{nyy}), we check the other indexes $i$ and $j$ by the  recurrence equation. For instance,
setting $i=1$ and $j=n_\beta-2$, we have
\begin{eqnarray*}
(\lambda_\alpha-\lambda_\beta)\langle\langle E_\beta^{(n_\beta-2)}|& \hat O&|D_\alpha^{(1)}\rangle\rangle
+\langle\langle E_\beta^{(n_\beta-2)}|\hat O|D_\alpha^{(0)}\rangle\rangle\nonumber\\
&&-\langle\langle E_\beta^{(n_\beta-1)} |\hat O|D_\alpha^{(1)}\rangle\rangle=0,
\end{eqnarray*}
It yields
\begin{eqnarray}
\langle\langle E_\beta^{(n_\beta-2)}| \hat O|D_\alpha^{(1)}\rangle\rangle=0,
\end{eqnarray}
by using Eqs. (\ref{nxx}) and (\ref{nyy}). As a result, when all of the indexes $i$ and $j$ are iterated, we conclude that
\begin{eqnarray}
\langle\langle E_\beta^{(j)}| \hat O|D_\alpha^{(i)}\rangle\rangle=0,\,\,\forall\,i,\,j,\label{a12}
\end{eqnarray}
for $\lambda_\alpha\neq\lambda_\beta$ for $\forall\, \alpha\neq\beta$.

In case of $\alpha=\beta$, Eq.(\ref{lambdat}) can be written as
\begin{eqnarray*}
\partial_t\lambda_\alpha\delta_{ij}=\langle\langle E_\alpha^{(j)}|\hat O|D_\alpha^{(i-1)}\rangle\rangle-
\langle\langle E_\alpha^{(j+1)}|\hat O|D_\alpha^{(i)}\rangle\rangle,
\end{eqnarray*}
Therefore, we can obtain the dynamical equation of the eigenvalues by taking sum over all indexes $i$ and $j$ of the Jordan block $\alpha$,
\begin{eqnarray*}
\partial_t\lambda_\alpha=\frac{1}{n_\alpha}\sum_{i,j=0}^{n_\alpha-1}\left(\langle\langle E_\alpha^{(j)}
|\hat O|D_\alpha^{(i-1)}\rangle\rangle-\langle\langle E_\alpha^{(j+1)}|\hat O|D_\alpha^{(i)}\rangle\rangle\right),
\end{eqnarray*}
If $i\neq j$, we find that
\begin{eqnarray*}
\langle\langle E_\alpha^{(j)}|\hat O|D_\alpha^{(i-1)}\rangle\rangle=\langle\langle E_\alpha^{(j+1)}
|\hat O|D_\alpha^{(i)}\rangle\rangle,
\end{eqnarray*}
Thus, we can write the dynamical equation of eigenvalues as
\begin{eqnarray*}
&\partial_t&\lambda_\alpha=\nonumber\\
&&\frac{1}{n_\alpha}\left(\sum_{i=0}^{n_\alpha-1}\langle\langle E_\alpha^{(i)}
|\hat O|D_\alpha^{(i-1)}\rangle\rangle-\sum_{j=0}^{n_\alpha-1}\langle\langle E_\alpha^{(j+1)}|\hat O
|D_\alpha^{(j)}\rangle\rangle\right),
\end{eqnarray*}
Replacing the index $j$ by $k=j+1$, it yields
\begin{eqnarray*}
&\partial_t&\lambda_\alpha\nonumber\\
&=&\frac{1}{n_\alpha}\left(\sum_{i=0}^{n_\alpha-1}\langle\langle E_\alpha^{(i)}
|\hat O|D_\alpha^{(i-1)}\rangle\rangle-\sum_{k=1}^{n_\alpha}\langle\langle E_\alpha^{(k)}|\hat O|D_\alpha^{(k-1)}
\rangle\rangle\right)\nonumber\\
&=&\frac{1}{n_\alpha}\left(\langle\langle E_\alpha^{(0)}|\hat O|D_\alpha^{(-1)}\rangle\rangle-\langle\langle
E_\alpha^{(n_\alpha)}|\hat O|D_\alpha^{(n_\alpha-1)}\rangle\rangle\right),
\end{eqnarray*}
Then, by considering Eqs. (\ref{rmax}) and (\ref{lmax}) , we finally obtain
\begin{eqnarray}
\partial_t\lambda_\alpha=0,
\end{eqnarray}
which implies that the dynamical invariants have indeed time-independent eigenvalues.

Let us consider now the solution of the master equation with the Liouvillian $\hat{ \mathcal L}_c(t)$, i.e.,
\begin{eqnarray}
\partial_t|\rho(t)\rangle\rangle=\hat{ \mathcal L}_c(t)|\rho(t)\rangle\rangle.\label{mel}
\end{eqnarray}
We expand the density matrix vector by the left basis vectors of the dynamical invariant $\hat{ \mathcal I }(t)$,
\begin{eqnarray}
|\rho(t)\rangle\rangle=\sum_{\alpha=0}^{m-1} c_\alpha(t)|\Phi_\alpha(t)\rangle\rangle,\label{dmv}
\end{eqnarray}
with
\begin{eqnarray}
|\Phi_\alpha(t)\rangle\rangle=\sum_{i=0}^{n_\alpha-1} b^\alpha_i(t)|D_\alpha^{(i)}(t)\rangle\rangle,\label{jex1}
\end{eqnarray}
where $m$ is the number of Jordan blocks. Inserting Eq.(\ref{dmv}) into Eq.(\ref{mel}), it yields
\begin{eqnarray}
\sum_{\alpha=0}^{m-1} \partial_tc_\alpha(t)|\Phi_\alpha(t)\rangle\rangle&+&\sum_{\alpha=0}^{m-1}
 c_\alpha(t)\partial_t|\Phi_\alpha(t)\rangle\rangle\nonumber\\
&=&\sum_{\alpha=0}^{m-1} c_\alpha(t)\hat{ \mathcal L}_c(t)|\Phi_\alpha(t)\rangle\rangle.\label{effeq1}
\end{eqnarray}
Here we define a left vector $\langle\langle\Psi_\beta(t)|$ of the Jordan block $\beta$, which satisfy
$\langle\langle\Psi_\beta(t)|\Phi_\alpha(t)\rangle\rangle=\delta_{\alpha\beta}$. This left vector can be
expanded by the left basis vectors of the Jordan block $\beta$, i.e., $\langle\langle\Psi_\beta(t)|=
\sum_{j=0}^{n_\beta-1}a_j^{\beta}(t)\langle\langle E_\beta^{(j)}(t)|$. Projecting Eq.(\ref{effeq1}) in
$\langle\langle\Psi_\beta(t)|$, we obtain
\begin{eqnarray}
\partial_t c_\beta(t)=\sum_{\alpha=0}^{m-1} c_\alpha(t)\langle\langle\Psi_\beta(t)|\hat{ O}(t)
|\Phi_\alpha(t)\rangle\rangle,\label{effeq2}
\end{eqnarray}
with $$\langle\langle\Psi_\beta(t)|\hat{ O}(t)|\Phi_\alpha(t)\rangle\rangle=\sum_{i,j}a_j^{\beta}(t)b^\alpha_i(t)
\langle\langle E_\beta^{(j)}| \hat O|D_\alpha^{(i)}\rangle\rangle.$$ By making use of Eq.(\ref{a12}), we have
\begin{eqnarray}
\partial_t c_\beta(t)= c_\beta(t)\langle\langle\Psi_\beta(t)|\hat{ O}(t)
|\Phi_\beta(t)\rangle\rangle.\label{effeq3}
\end{eqnarray}
This results in a formal solution of the density matrix vector
\begin{eqnarray}
|\rho(t)\rangle\rangle=\sum_{\alpha=0}^{m-1} c_\alpha(0)|\tilde\Phi_\alpha(t)\rangle\rangle,\label{fs1}
\end{eqnarray}
with ``dynamical modes'' $|\tilde\Phi_\alpha(t)\rangle\rangle=\exp(\eta_\alpha(t))|\Phi_\alpha(t)\rangle\rangle$,
where the phases are defined as
\begin{eqnarray}
\eta_\alpha(t)=\int_0^t d\tau \langle\langle\Psi_\alpha(\tau)|\hat{ O}(\tau)|\Phi_\alpha(\tau)\rangle\rangle.
\end{eqnarray}

In fact, we do not obtain a complete solution of the master equation Eq.(\ref{mel}), since the coefficients in
$|\phi_\alpha(t)\rangle\rangle$ have not been determined. This will be an interesting and open question for further\
investigation. But the formal solution in Eq.(\ref{fs1}) is enough to establish the shortcuts to adiabaticity of
open quantum systems, since the adiabatic theorem of open quantum systems just requires that the transition
between different Jordan blocks are forbidden \cite{Sarandy2005}. Putting undetermined  coefficients in $|\phi_\alpha(t)\rangle\rangle$
aside, we only need to ensure the quantum state in the same Jordan block of $\hat{ \mathcal L}_c(t)$ at the beginning and end of
control process.

\section{Comparison with the Transitionless Driving Scheme of Open Quantum Systems}\label{AB}

In the following, we present a proof  that our scheme is as general as, in some cases it is beyond, the G. Vacanti's method \cite{Vacanti2014}.

For the reference Liouvillian superoperator $\hat{ \mathcal L}_0(t)$, it is always possible to find a similarity transformation $C(t)$ such that
$\hat {\mathcal L}_0(t)$ is written in the canonical Jordan form
 \begin{eqnarray}
\hat{\mathcal L}_J(t)=C^{-1}(t)\hat {\mathcal L}_0(t)C(t)=\text{diag}[J_1(t),...,J_N(t)],\label{lj}
 \end{eqnarray}
 where $J_\alpha(t )$ represents the Jordan block (of dimension $n_\alpha$) corresponding to the eigenvalue $\zeta_\alpha(t)$
 of $\hat {\mathcal L}_0(t)$. The number $N$ of Jordan blocks is equal to the number of linear independent eigenvectors of $\hat {\mathcal L}_0(t)$ and the
 similarity transformation is given by
 \begin{eqnarray*}
C(t)=\sum_{\alpha=1}^{N}\sum_{i=1}^{n_\alpha} |D_\alpha^{(i)}(t)\rangle\rangle\langle\langle \sigma_\alpha^{(i)}|,
 \end{eqnarray*}
 where $\{|D_\alpha^{(i)}(t)\rangle\rangle\}$ is a right instantaneous quasi-eigenbases of $\hat {\mathcal L}_0(t)$  associated
with the eigenvalues $\{\zeta_\alpha(t)\}$ which satisfies
  \begin{eqnarray*}
\hat {\mathcal L}_0\,|D_\alpha^{(i)}\rangle\rangle=\zeta_\alpha|D_\alpha^{(i)}\rangle\rangle+|D_\alpha^{(i-1)}\rangle\rangle.
\end{eqnarray*}
And $\{ |\sigma_\alpha^{(i)}\rangle\rangle\}$ is a set of time-independent bases which is used to calculate  the matrix form of $\hat {\mathcal L}_0(t)$.

Here we construct a dynamical invariant superoperator $\hat {\mathcal I}(t)$ which have same Jordan blocks structure. In other words,
 $\hat {\mathcal I}(t)$ can be diagonal by the same similarity transformation $C(t)$, i.e.,
  \begin{eqnarray}
\hat{\mathcal I}_J=C^{-1}(t)\hat{ \mathcal I}(t)C(t)=\text{diag}[I_1,...,I_N],\label{Ij}
 \end{eqnarray}
where $I_\alpha$ is the Jordan block (of dimension $n_\alpha$) corresponding to the eigenvalue $\lambda_\alpha$  of $\hat {\mathcal I}(t)$.
The eigenvalues $\{\lambda_\alpha\}$ are time-independent. In fact,
we choose a dynamical invariant with the same Jordan blocks structure is equivalent to choose the trajectory of the mixed-state inverse
engineering as the adiabatic trajectory. Therefore, $\hat {\mathcal L}_0(t)$ and ${\hat{ \mathcal I}}(t)$ share common quasi-eigenvectors
$|D_{\alpha}^{(i)}(t)\rangle\rangle$, i.e.,
  \begin{eqnarray*}
\hat {\mathcal I}\,|D_\alpha^{(i)}\rangle\rangle=\lambda_\alpha|D_\alpha^{(i)}\rangle\rangle+|D_\alpha^{(i-1)}\rangle\rangle.
\end{eqnarray*}

The dynamical invariants $\hat{\mathcal I}(t)$  satisfy the dynamical equation Eq.(\ref{a10})
Following the G. Vacanti method, we set that the control Liouvillian superoperator
can be written as $\hat{\mathcal L}(t)=\hat{\mathcal L}_0(t)+\hat{\mathcal L}_c(t)$, where $\hat{\mathcal L}_c(t)$ is the counterdiabatic
superoperator. Substituting Eq.(\ref{Ij}) into Eq.(\ref{a10}), it yields
\begin{eqnarray}
\partial_{t}\left(C\hat{\mathcal{I}}_{J}C^{-1}\right)&=&[C\hat{\mathcal{L}}_{J}C^{-1}+\hat{\mathcal{L}}_{c},C\hat{\mathcal{I}}_{J}C^{-1}]
\end{eqnarray}
where Eq.(\ref{lj}) has been used. Considering that $\hat{\mathcal{I}}_{J}$ is time-independent and taking the same similarity transformation
$C(t)$ on above equation, we obtain
\begin{eqnarray}
\left[C^{-1}\partial_{t}C,\hat{\mathcal{I}}_{J}\right]=\left[\hat{\mathcal{L}}_{J},\hat{\mathcal{I}}_{J}\right]
+\left[C^{-1}\,\hat{\mathcal{L}}_{c}C,\hat{\mathcal{I}}_{J}\right].\label{a11}
\end{eqnarray}
in which the fact $C^{-1}\partial_{t}C=-\partial_{t}C^{-1}\,C$ are considered. Since $\hat{\mathcal I}(t)$ and $\hat{\mathcal L}_0(t)$ have
same Jordan blocks structure, we have $ [\hat{\mathcal{L}}_{J},\hat{\mathcal{I}}_{J}]=0$. It is not difficult to see that,
if $ \hat{\mathcal{L'}}_{c}\equiv C^{-1}\,\hat{\mathcal{L}}_{c}C=C^{-1}\partial_{t}C$, Eq.(\ref{a11}) holds. We expand $ \hat{\mathcal{L'}}_{c}$
by the bases $\{ |\sigma_\alpha^{(i)}\rangle\rangle\}$, and separate it into two parts, $ \hat{\mathcal{L'}}_{c}= \hat{\mathcal{L'}}_{J}+
 \hat{\mathcal{L'}}_{nd}$, where
\begin{eqnarray}
 &&\hat{\mathcal{L'}}_{J}=\sum_{\alpha,i,j}C_\alpha^{i,j} |\sigma_\alpha^{(i)}\rangle\rangle\langle\langle\sigma_\alpha^{(j)}|,\nonumber\\
 &&\hat{\mathcal{L'}}_{nd}=\sum_{\alpha\neq\beta,i,j}C_{\alpha,\beta}^{i,j} |\sigma_\alpha^{(i)}\rangle\rangle\langle\langle\sigma_\beta^{(j)}|,\nonumber
\end{eqnarray}
with $C_{\alpha,\beta}^{i,j}=\langle\langle\sigma_\alpha^{(j)}|C^{-1}\partial_{t}C|\sigma_\beta^{(i)}\rangle\rangle$. The superoperator
$\hat{\mathcal{L'}}_{nd}$ is used to forbid the transitions from $J_\alpha$ to $J_\beta$. Therefore, the counterdiabatic superoperator which
really required for STAs is
 \begin{eqnarray}
\hat{\mathcal{L}}_{\text{tqd}}(t)=C(t)\hat{\mathcal{L'}}_{nd}(t)C^{-1}(t).\label{tqd}
\end{eqnarray}
Comparing Eq.(\ref{tqd}) with Eq. (15)  in Ref.\cite{Vacanti2014}, we immediately find that $\hat{\mathcal{L}}_{\text{tqd}}(t)$
is the very counterdiabatic superoperator given by  G. Vacanti and his co-authors \cite{Vacanti2014}.

Above simple proof illustrates that, if the trajectory of the general STAs based on the invariant theory of open quantum systems is chosen as the adiabatic trajectory,
our method is coincident with the transitionless quantum driving method proposed in Ref.\cite{Vacanti2014}.
However, the adiabatic trajectory is not the only choice of the trajectories in our scheme. There are many trajectories can be used
to inversely engineer the open quantum system. As shown by the example in Sec. \ref{secIIIB},  proper
trajectories can always produce reasonable and applicable control protocols, which helps us to overcome the difficulties met in the
control of   microscopic or/and mesoscopic systems. Hence,  the mixed-state inverse engineering is more general than the
transitionless quantum driving method of open quantum systems proposed by G. Vacanti \cite{Vacanti2014}.

The main difficulty in the G. Vacanti's method is how to realize the counterdiabatic
superoperator into a practical control. The MIE scheme solves those problems by selecting  proper trajectories. According to the
symmetry of the open quantum systems, we can choose flexibly  the form of the control Liouvillian, this make the
 proposal widely applicable in the control of open quantum systems.

 \section{The Instantaneous Steady State of $\hat{\mathcal L}_0(t)$} \label{AC}

 We use the ``bra-ket'' notation for the superoperator to rewrite the master equation Eq.(\ref{me3}),
and reshape the density matrix into a $1\times 9$ complex vector. The density matrix vector can be written as $$\ket{\rho}\rangle=
(\rho_{-1\,-1},\rho_{-1\,2},\rho_{-1\,1},\rho_{2\,-1},\rho_{2\,2},\rho_{2\,1},\rho_{1\,-1},\rho_{1\,2},\rho_{1\,1})^{\text{T}}$$ with $\rho_{i\, j}=\bra{i}\hat \rho\ket{j}$.
In order to present an analytic result, we assume that  $\Gamma_{-1}=\Gamma_{+1}\equiv\Gamma$ and $\omega_{2\rightarrow-1}=\omega_{2\rightarrow+1}
\equiv\omega_0$. At room temperature (T=300 K), the mean excitation number is $N=1.9\times10^{-33}$. For practice application, we choose the instantaneous steady state of the reference Liouvillian $\hat{ \mathcal L_0}$ as the trajectory
of inverse engineering, in this case  the  dephasing is the key obstacle for the performance of the protocol. The reference
Liouvillian superoperator $\hat{ \mathcal L_0}$ can be expressed as a $9\times 9$ matrix,
\begin{widetext}
\begin{equation}
\hat{ \mathcal L_0}=\hbar\Gamma\left(\begin{array}{ccccccccc} - 2\, \mathrm{N}\,  & \mathrm{i}\mathrm{\Omega_p/\Gamma} & 0 & -\mathrm{i} \mathrm{\Omega_p/\Gamma} &
2\, \, \left(\mathrm{N} + 1\right) & 0 & 0 & 0 & 0\\ \mathrm{i} \mathrm{\Omega_p/\Gamma}& - \, \left(3\, \mathrm{N} + 2\right) &
\mathrm{i}\mathrm{\Omega_s/\Gamma} & 0 & - \mathrm{i} \mathrm{\Omega_p/\Gamma}& 0 & 0 & 0 & 0\\ 0 & \mathrm{i} \mathrm{\Omega_s/\Gamma}&
- 2\, \mathrm{N}\,  & 0 & 0 & - \mathrm{i}\mathrm{\Omega_p/\Gamma} & 0 & 0 & 0\\ - \mathrm{i} \mathrm{\Omega_p/\Gamma} & 0
& 0 & - \, \left(3\, \mathrm{N} + 2\right) &\mathrm{i} \mathrm{\Omega_p/\Gamma} & 0 & - \mathrm{i}\mathrm{\Omega_s/\Gamma} &
0 & 0\\ 2\, \mathrm{N}\,  & - \mathrm{i} \mathrm{\Omega_p/\Gamma}& 0 &\mathrm{i} \mathrm{\Omega_p/\Gamma} &
- 4\, \, \left(\mathrm{N} + 1\right) & \mathrm{i}\mathrm{\Omega_s/\Gamma} & 0 & - \mathrm{i}\mathrm{\Omega_s/\Gamma} &
2\, \mathrm{N}\, \\ 0 & 0 & - \mathrm{i} \mathrm{\Omega_p/\Gamma} & 0 &\mathrm{i} \mathrm{\Omega_s/\Gamma} &
- \, \left(3\, \mathrm{N} + 2\right) & 0 & 0 & - \mathrm{i}\mathrm{\Omega_s/\Gamma}\\ 0 & 0 & 0 &
-\mathrm{i} \mathrm{\Omega_s/\Gamma}& 0 & 0 & - 2\, \mathrm{N}\,  &\mathrm{i} \mathrm{\Omega_p/\Gamma}
& 0\\ 0 & 0 & 0 & 0 & -\mathrm{i} \,\mathrm{\Omega_s/\Gamma}  & 0 & \mathrm{i}\mathrm{\Omega_p/\Gamma} &
- \, \left(3\, \mathrm{N} + 2\right) &\mathrm{i} \mathrm{\Omega_s/\Gamma}\\ 0 & 0 & 0 & 0
& 2\, \, \left(\mathrm{N} + 1\right) & -\mathrm{i} \mathrm{\Omega_s/\Gamma} & 0 &
\mathrm{i} \mathrm{\Omega_s/\Gamma}& - 2\, \mathrm{N}\,  \end{array}\right).\nonumber
\end{equation}
The steady state is obtained immediately by considering $\hat{ \mathcal L_0}\ket{\rho_0}\rangle=0$,
\begin{equation}
\ket{\rho_0}\rangle=\frac{1}{z}\left(\begin{array}{ccccccccc}
{\left(3\, \mathrm{N}^2 + 2\mathrm{N}\right)\, \left(\mathrm{N} + 1\right)\, {\mathrm{\Gamma}}^2 + \mathrm{N}\, {\mathrm{\Omega}}^2 +  {\mathrm{\Omega_p}}^2}\\
{-{i}\, \mathrm{N}\, \mathrm{\Gamma}\, \mathrm{\Omega_p}}\\
{-\mathrm{\Omega_p}\, \mathrm{\Omega_s}}\\
{{i}\, \mathrm{N}\, \mathrm{\Gamma}\, \mathrm{\Omega_p}}\\
{\mathrm{N}^2\, \left(3\, \mathrm{N} + 2\right)\, {\mathrm{\Gamma}}^2 + \mathrm{N}\, {\mathrm{\Omega}}^2 }\\
{{i}\, \mathrm{N}\, \mathrm{\Gamma}\, \mathrm{\Omega_s}}\\
{-\mathrm{\Omega_p}\, \mathrm{\Omega_s}}\\
{-{i}\, \mathrm{N}\, \mathrm{\Gamma}\, \mathrm{\Omega_s}}\\
{\left(3\, \mathrm{N}^2 + 2\mathrm{N}\right)\, \left(\mathrm{N} + 1\right)\, {\mathrm{\Gamma}}^2 + \mathrm{N}\, {\mathrm{\Omega}}^2 +  {\mathrm{\Omega_s}}^2}\\
\end{array}\right).\label{is3}
\end{equation}
with the normalized factor $$z={\left(3\, \mathrm{N} + 1\right)\, {\mathrm{\Omega}}^2 + \mathrm{N}\, {\mathrm{\Gamma}}^2\, {\left(3\, \mathrm{N} + 2\right)}^2}.$$
For $N=0$, the instantaneous steady state is the dark state of the Hamiltonian Eq.(\ref{rh3}), i.e. $\ket{\rho_0}\rangle=(\cos^2\theta,0,\sin\theta\cos\theta,0,0,0,
\sin\theta\cos\theta,0,\sin^2\theta)^{\text{T}}$.

We parameterize the adiabatic trajectory given by the instantaneous steady state of $\hat{ \mathcal L_0}$ via the generalized Bloch vector $\{r_k\}_{k=1}^8$, which expands the density matrix of the
three-level system as follows,
\begin{equation}
\rho(t)=\frac{1}{3}\left(\text{I}+\sqrt{3}\sum_{k=1}^8 r_k(t) T_k\right),
\end{equation}
where $\text{I}$ is a $3\times 3$ identity matrix, and $T_k$ denotes the regular Gellmann matrix
\begin{eqnarray}
&T_1=\left(\begin{array}{ccc}
0 & 1 & 0\\
1 & 0 & 0\\
0 & 0 & 0
\end{array}\right),\,
T_2=\left(\begin{array}{ccc}
0 & -i & 0\\
i & 0 & 0\\
0 & 0 & 0
\end{array}\right),\,
T_3=\left(\begin{array}{ccc}
1 & 0 & 0\\
0 & -1 & 0\\
0 & 0 & 0
\end{array}\right),\,
T_4=\left(\begin{array}{ccc}
0 & 0 & 1\\
0 & 0 & 0\\
1 & 0 & 0
\end{array}\right),\,\nonumber\\
&T_5=\left(\begin{array}{ccc}
0 & 0 & -i\\
0 & 0 & 0\\
i & 0 & 0
\end{array}\right),\,
T_6=\left(\begin{array}{ccc}
0 & 0 & 0\\
0 & 0 & 1\\
0 & 1 & 0
\end{array}\right),\,
T_7=\left(\begin{array}{ccc}
0 & 0 & 0\\
0 & 0 & -i\\
0 & i & 0
\end{array}\right),\,
T_8=\frac{1}{\sqrt{3}}\left(\begin{array}{ccc}
1 & 0 & 0\\
0 & 1 & 0\\
0 & 0 & -2
\end{array}\right).\label{Tx}
\end{eqnarray}
These $\{T_\mu\}$ span all traceless Hermitian matrices of the Lie algebra su(3). Thus, the Bloch vectors corresponding to
 the instantaneous steady state
Eq.(\ref{is3}) are
\begin{eqnarray}
&&r_2=\sqrt{3}\,\mathrm{N}\, \mathrm{\Gamma}\, \mathrm{\Omega_p}/z,\nonumber\\
&&r_3=\sqrt{3}\,\left(\left(3\, \mathrm{N}^2 + 2\mathrm{N}\right)\, {\mathrm{\Gamma}}^2 +  {\mathrm{\Omega_s}}^2 \right)/(2z),\nonumber\\
&&r_4=-\sqrt{3}\,\mathrm{\Omega_p}\, \mathrm{\Omega_s}/z,\nonumber\\
&&r_7=-\sqrt{3}\,\mathrm{N}\, \mathrm{\Gamma}\, \mathrm{\Omega_s}/z,\nonumber\\
&&r_8=-\left(\left(3\, \mathrm{N}^2 + 2\mathrm{N}\right)\, {\mathrm{\Gamma}}^2+2\,{\mathrm{\Omega_p}}^2 -{\mathrm{\Omega_s}}^2 \right)/(2z),\label{r38}
\end{eqnarray}
and the other components are zeros. Correspondingly, the dynamical invariants can be parameterized by the Bloch vector according to our proposal (see Eq.(\ref{di11})),
\begin{eqnarray}
\hat{\mathcal{I}}(t)=\frac{\sqrt{3}}{3}\,\Omega_{I}\left(\begin{array}{ccccccccc}
\left(\mathrm{r_{3}}+\frac{\sqrt{3}\,\mathrm{r_{8}}}{3}\right)+\frac{\sqrt{3}}{3} & 0 & 0 & 0 & \left(\mathrm{r_{3}}+\frac{\sqrt{3}\,\mathrm{r_{8}}}{3}\right)+\frac{\sqrt{3}}{3} & 0 & 0 & 0 & \left(\mathrm{r_{3}}+\frac{\sqrt{3}\,\mathrm{r_{8}}}{3}\right)+\frac{\sqrt{3}}{3}\\
\mathrm{i}\,\mathrm{r_{2}} & 0 & 0 & 0 & \mathrm{i}\,\mathrm{r_{2}} & 0 & 0 & 0 & \mathrm{i}\,\mathrm{r_{2}}\\
\mathrm{r_{4}} & 0 & 0 & 0 & \mathrm{r_{4}} & 0 & 0 & 0 & \mathrm{r_{4}}\\
-\mathrm{i}\,\mathrm{r_{2}} & 0 & 0 & 0 & -\mathrm{i}\,\mathrm{r_{2}} & 0 & 0 & 0 & -\mathrm{i}\,\mathrm{r_{2}}\\
\frac{\sqrt{3}}{3}\,-\left(\mathrm{r_{3}}-\frac{\sqrt{3}\,\mathrm{r_{8}}}{3}\right) & 0 & 0 & 0 & \frac{\sqrt{3}}{3}\,-\left(\mathrm{r_{3}}-\frac{\sqrt{3}\,\mathrm{r_{8}}}{3}\right) & 0 & 0 & 0 & \frac{\sqrt{3}}{3}\,-\left(\mathrm{r_{3}}-\frac{\sqrt{3}\,\mathrm{r_{8}}}{3}\right)\\
\mathrm{i}\,\mathrm{r_{7}} & 0 & 0 & 0 & \mathrm{i}\,\mathrm{r_{7}} & 0 & 0 & 0 & \mathrm{i}\,\mathrm{r_{7}}\\
\mathrm{r_{4}} & 0 & 0 & 0 & \mathrm{r_{4}} & 0 & 0 & 0 & \mathrm{r_{4}}\\
-\mathrm{i}\,\mathrm{r_{7}} & 0 & 0 & 0 & -\mathrm{i}\,\mathrm{r_{7}} & 0 & 0 & 0 & -\mathrm{i}\,\mathrm{r_{7}}\\
\frac{\sqrt{3}}{3}-\frac{2\,\sqrt{3}\,\mathrm{r_{8}}}{3} & 0 & 0 & 0 & \frac{\sqrt{3}}{3}-\frac{2\,\sqrt{3}\,\mathrm{r_{8}}}{3} & 0 & 0 & 0 & \frac{\sqrt{3}}{3}-\frac{2\,\sqrt{3}\,\mathrm{r_{8}}}{3}
\end{array}\right)
\end{eqnarray}

where $\Omega_I$ is an arbitrary nonzero constant.\

\section{The Control Parameters in the Control Liouvillian} \label{AD}

Considering the dynamical equation of the dynamical invariants Eq.(\ref{a10}), we can determine all of control parameters in the control Liouvillian $\hat{ \mathcal L_c}$,
\begin{eqnarray}
\Omega_{s}^i=n_{s}/d,\,\Omega_{p}^i=n_{p}/d,\,\Omega_{c}^i=n_{c}/d,\,N_{-1}^i=n_{-1}/d,\,N_{+1}^i=n_{+1}/d, \label{cp3}
\end{eqnarray}
in which
\begin{eqnarray*}
d&=&2\,\sqrt{3}\,\mathrm{r_4}^{4}\,C_{2}+4\,\sqrt{3}\,\mathrm{r_2}^{4}\,C_{1}+\mathrm{r_2}^{2}\,\left(12\,\mathrm{r_8}\,\left(\mathrm{r_4}^{2}+\mathrm{r_7}^{2}\right)+12\,\sqrt{3}\,\left(\mathrm{r_3^{2}}+\mathrm{r_8^{2}}\right)\,C_{1}\right)\\
&&+\mathrm{r_4}^{2}\,\left(6\,\mathrm{r_7}^{2}\,C_{2}-8\,\sqrt{3}\,\mathrm{r_3}\,C_{1}C_{2}\right)+8\,\mathrm{r_3}^{2}\,C_{2}\,C_{1}^{2}+4\,\sqrt{3}\,\mathrm{r_3}\,\mathrm{r_7}^{2}\,\left(\mathrm{r_2}^{2}-2\,\mathrm{r_7}^{2}\right)\\
&&-8\,\mathrm{r_2}\,\mathrm{r_4}\,\mathrm{r_7}\,\left(\sqrt{3}\,\mathrm{r_7}^{2}+3\,C_{2}\,C_{4}\right)+8\,\sqrt{3}\,\mathrm{r_2}^{3}\,\mathrm{r_4}\,\mathrm{r_7}+12\,\mathrm{r_3}\,\mathrm{r_7}^{2}\,\left(\mathrm{r_3}-\left(\sqrt{3}-2\right)\,\mathrm{r_8}\right)\,\left(\left(2\,\sqrt{3}+3\right)\,\mathrm{r_8}+\sqrt{3}\,\mathrm{r_3}\right),
\end{eqnarray*}
\begin{eqnarray*}
n_{s}&=&2\,\sqrt{3}\,\partial_{t}r_{2}\left(\mathrm{r_{4}}^{2}\,\left(\mathrm{r_{7}}^{2}-C_{1}C_{2}\right)+2\,\mathrm{r_{2}}^{2}\,\mathrm{r_{3}}\,C_{1}+\mathrm{r_{2}}\,\mathrm{r_{4}}\,\mathrm{r_{7}}\,\left(C_{2}+4\,\mathrm{r_{3}}\right)+2\,\mathrm{r_{3}}\,C_{2}\,\left(2\,\mathrm{r_{7}}^{2}+C_{1}^{2}\right)\right)\nonumber\\
&&+\sqrt{3}\,\partial_{t}r_{3}\left(-2\,\mathrm{r_{2}}^{3}\,C_{1}+2\,\mathrm{r_{2}}\,\left(-C_{1}^{2}C_{2}-2\,\mathrm{r_{7}}^{2}\,\left(C_{2}+\mathrm{r_{3}}\right)\right)-\mathrm{r_{4}}\,\mathrm{r_{7}}\,\left(4\,\mathrm{r_{2}}^{2}+\mathrm{r_{4}}^{2}-\mathrm{r_{7}}^{2}+2\,\sqrt{3}\,\mathrm{r_{8}}\,C_{1}\right)\right)\\
&&+\partial_{t}r_{4}\left(-\mathrm{r_{7}}\,\left(4\,\mathrm{r_{3}}\,C_{1}^{2}-6\,\mathrm{r_{4}}^{2}\,C_{3}\right)-4\,\sqrt{3}\,\mathrm{r_{3}}\,\mathrm{r_{7}}\,\left(\mathrm{r_{2}}^{2}+2\,\mathrm{r_{7}}^{2}\right)-2\,\sqrt{3}\,\mathrm{r_{2}}\,\mathrm{r_{4}}\,\left(2\,C_{1}C_{2}+3\,\mathrm{r_{7}}^{2}\right)\right)\\&&
+2\,\sqrt{3}\,\partial_{t}r_{7}\left(2\,\mathrm{r_{3}}\,\mathrm{r_{4}}\,C_{1}C_{2}-\mathrm{r_{4}}^{3}\,C_{2}-\left(2\,\sqrt{3}\,\mathrm{r_{2}}\,\mathrm{r_{8}}-\mathrm{r_{4}}\,\mathrm{r_{7}}\right)\,\left(\mathrm{r_{2}}\,\mathrm{r_{4}}+2\,\mathrm{r_{3}}\,\mathrm{r_{7}}\right)\right)\\
&&+\partial_{t}r_{8}\,\left(\mathrm{r_{2}}\,\left(2\,C_{1}C_{2}^{2}+2\,\mathrm{r_{4}}^{2}\,C_{2}+2\,\mathrm{r_{7}}^{2}\,\left(\mathrm{r_{3}}+C_{1}\right)\right)+2\,\mathrm{r_{2}}^{3}\,C_{2}+\mathrm{r_{4}}\,\mathrm{r_{7}}\,\left(4\,C_{2}^{2}+2\,\mathrm{r_{2}}^{2}+\mathrm{r_{4}}^{2}+\mathrm{r_{7}}^{2}+2\,\sqrt{3}\,\mathrm{r_{8}}\,C_{1}\right)\right)\\
&&+\Gamma\,\mathrm{r_{2}}^{2}\,\left(14\,\mathrm{r_{4}}\,\mathrm{r_{7}}+2\,\mathrm{r_{4}}\,\mathrm{r_{7}}\,\left(C_{3}+3\,\sqrt{3}\,\mathrm{r_{3}}\right)\right)-\mathrm{r_{7}}^{3}\,\left(2\,\mathrm{r_{4}}+2\,\mathrm{r_{4}}\,\left(C_{3}-6\,\sqrt{3}\,\mathrm{r_{3}}\right)\right)\\
&&-\Gamma\,\mathrm{r_{7}}\,\left(\mathrm{r_{4}}^{3}\,\left(8\,C_{3}-4\right)-\mathrm{r_{4}}\,\left(4\,\sqrt{3}\,\mathrm{r_{3}}^{3}-44\,\mathrm{r_{3}}^{2}\,\mathrm{r_{8}}+4\,\mathrm{r_{3}}^{2}+4\,\sqrt{3}\,\mathrm{r_{3}}\,\mathrm{r_{8}}^{2}+16\,\sqrt{3}\,\mathrm{r_{3}}\,\mathrm{r_{8}}-12\,\mathrm{r_{8}}^{3}-12\,\mathrm{r_{8}}^{2}\right)\right)\\
&&-\Gamma\,\mathrm{r_{2}}^{3}\,\left(4\,\sqrt{3}\,\mathrm{r_{8}}-8\,\mathrm{r_{3}}+4\,\mathrm{r_{8}}\,C_{2}\right)+\Gamma\,\mathrm{r_{2}}\,\mathrm{r_{7}}^{2}\,\left(16\,\mathrm{r_{3}}+4\,\sqrt{3}\,\mathrm{r_{8}}+4\,\mathrm{r_{8}}\,C_{2}\right)\\
&&+\Gamma\,\mathrm{r_{2}}\,\mathrm{r_{4}}^{2}\,\left(8\,\sqrt{3}\,\mathrm{r_{7}}^{2}+\left(2-6\,\sqrt{3}\mathrm{r_{3}}+14\,\mathrm{r_{8}}\right)\,C_{2}\right)-\Gamma\,\mathrm{r_{2}}\,C_{1}C_{2}\,\left(4\,\sqrt{3}\,\mathrm{r_{8}}-8\,\mathrm{r_{3}}+4\,\mathrm{r_{8}}\,C_{2}\right),
\end{eqnarray*}
\begin{eqnarray*}
n_{p}&=&\partial_{t}r_{2}\,\left(6\,\mathrm{r_4}^{3}\,C_{2}-\mathrm{r_4}\,\left(12\,\mathrm{r_3}\,C_{1}C_{2}-6\,\sqrt{3}\,\mathrm{r_7}^{2}\,C_{3}\right)+6\,\mathrm{r_2}\,\mathrm{r_7}\,\left(\mathrm{r_4}^{2}+\sqrt{3}\,C_{1}C_{3}\right)+6\,\mathrm{r_2}^{2}\,\mathrm{r_4}\,C_{1}\right)\\
&&+\partial_{t}r_{3}\,\left(3\,\mathrm{r_4}^{2}\,\mathrm{r_7}\,C_{2}-6\,\mathrm{r_2}^{2}\,\mathrm{r_7}\,C_{1}-3\,\mathrm{r_2}\,\mathrm{r_4}\,\left(\mathrm{r_4}^{2}+3\,\mathrm{r_7}^{2}-2\,C_{1}\,\left(\mathrm{r_3}+C_{2}\right)\right)+6\,\mathrm{r_7}\,-C_{1}\,\left(\mathrm{r_3}\,C_{2}-\frac{\mathrm{r_7}^{2}}{2}\right)\right)\\
&&+6\,\partial_{t}r_{4}\,\left(-2\,C_{1}\,\mathrm{r_2}^{3}-3\,\mathrm{r_4}\,\mathrm{r_7}\,\mathrm{r_2}^{2}+\left(-2\,\sqrt{3}\,\mathrm{r_8}\,\mathrm{r_4}^{2}-C_{1}\,\left(4\,\mathrm{r_3}^{2}+\mathrm{r_7}^{2}\right)\right)\,\mathrm{r_2}-4\,\mathrm{r_3}\,\mathrm{r_4}\,\mathrm{r_7}\,C_{2}\right)\\
&&+\partial_{t}r_{7}\,\left(\mathrm{r_2}^{2}\,\left(-12\,C_{1}C_{2}+6\,\mathrm{r_4}^{2}\right)-24\,\mathrm{r_3}^{2}\,C_{1}C_{2}+12\,\mathrm{r_3}\,\mathrm{r_4}^{2}\,C_{2}+12\,\mathrm{r_3}\,\mathrm{r_7}^{2}\,C_{1}+6\,\mathrm{r_2}\,\mathrm{r_4}\,\mathrm{r_7}\,\left(3\,C_{1}-2\,\mathrm{r_3}\right)\right)\\
&&+\partial_{t}r_{8}\,\left(\sqrt{3}\,\mathrm{r_4}^{2}\,\mathrm{r_7}\,C_{2}-4\,\sqrt{3}\,\mathrm{r_2}^{2}\,\mathrm{r_7}\,C_{2}-\sqrt{3}\,\mathrm{r_2}\,\mathrm{r_4}\,\left(2\,\left(\mathrm{r_3}+C_{2}\right)^{2}-5\,\mathrm{r_7}^{2}\right)-\sqrt{3}\,\mathrm{r_2}\,\mathrm{r_4}\,\left(2\,\mathrm{r_2}^{2}-\mathrm{r_4}^{2}\right)\right)\\
&&+\partial_{t}r_{8}\,\left(-7\,\sqrt{3}\,\mathrm{r_3}\,\mathrm{r_7}\,\left(2\,\mathrm{r_3}^{2}-\mathrm{r_7}^{2}\right)+2\,\sqrt{3}\,\mathrm{r_3}\,\mathrm{r_8}\,\left(\mathrm{r_2}\,\mathrm{r_4}-3\,\mathrm{r_7}\,\mathrm{r_8}\right)-3\,\mathrm{r_7}\,\mathrm{r_8}\,\left(4\,\mathrm{r_3}-\mathrm{r_7}\right)\,\left(4\,\mathrm{r_3}+\mathrm{r_7}\right)\right)\\
&&+\Gamma\,\mathrm{r_2}\,\left(2\,\mathrm{r_7}^{2}\,\left(-2\,\mathrm{r_3}+5\,C_{1}+7\,\sqrt{3}\right)-4\,\sqrt{3}\,\left(\mathrm{r_8}+1\right)\,\left(4\,\mathrm{r_3}^{2}+C_{1}C_{2}\right)+36\,\mathrm{r_3}\,\mathrm{r_4}\,\left(\mathrm{r_3}^{2}+\mathrm{r_8}^{2}\right)\right)\\
&&+\Gamma\,\left(-\mathrm{r_7}^{3}\,\left(-4\,\mathrm{r_8}-10\,C_{4}+2\,\sqrt{3}\,C_{2}C_{3}\right)+\mathrm{r_2}^{2}\,\left(2\,\mathrm{r_7}\,C_{2}^{2}-12\,\mathrm{r_7}\,\left(3\,\mathrm{r_8}-2\,\mathrm{r_4}^{2}\right)+2\,\mathrm{r_7}\,C_{2}\,\left(2\,\mathrm{r_3}+\sqrt{3}\right)\right)\right)\\
&&+\Gamma\,\left(\mathrm{r_2}^{3}\,\left(18\,\mathrm{r_4}\,C_{1}-2\,\sqrt{3}\,\mathrm{r_4}\,\left(2\,\mathrm{r_8}+1\right)\right)+4\,\sqrt{3}\,\left(4\,\mathrm{r_8}+1\right)\,\mathrm{r_2}\,\mathrm{r_4}^{3}\right)\\
&&-\Gamma\,\mathrm{r_7}\,\left(\mathrm{r_4}^{2}\,C_{2}\,\left(2\,\sqrt{3}-30\,\mathrm{r_3}+2\sqrt{3}\,\mathrm{r_8}\right)-4\,\mathrm{r_3}\,\left(C_{2}\,\left(2\,\mathrm{r_8}-5\,C_{4}\right)+\mathrm{r_3}\,C_{2}\,\left(5\,C_{3}-2\,\mathrm{r_3}\right)-\sqrt{3}\,\mathrm{r_8}\,C_{1}C_{2}\right)\right),
\end{eqnarray*}
\begin{eqnarray*}
n_{c}&=&\partial_{t}r_{2}\,\left(6\,\mathrm{r_2}\,\mathrm{r_4}\,\left(\mathrm{r_7}^{2}+\sqrt{3}\,C_{1}C_{4}\right)+12\,\mathrm{r_3}\,\mathrm{r_7}\,\left(\mathrm{r_4}^{2}+2\,\mathrm{r_7}^{2}\right)-6\,\mathrm{r_2}^{2}\,\mathrm{r_7}\,C_{1}+12\,\mathrm{r_3}\,\mathrm{r_7}\,C_{1}^{2}\right)\\
&&+\partial_{t}r_{3}\,\left(\mathrm{r_4}\,\left(6\,\mathrm{r_3}\,C_{1}^{2}+3\,\mathrm{r_7}^{2}\,C_{2}\right)-3\,\mathrm{r_4}^{3}\,C_{1}-6\,\mathrm{r_2}^{2}\,\mathrm{r_4}\,C_{1}-\mathrm{r_2}\,\mathrm{r_7}\,\left(3\,\mathrm{r_4}^{2}+9\,\mathrm{r_7}^{2}+6\,C-1\,\left(\mathrm{r_3}-C_{1}\right)\right)\right)\\
&&+\partial_{t}r_{4}\,\left(12\,\mathrm{r_3}\,\mathrm{r_4}^{2}\,C_{1}-\mathrm{r_2}^{2}\,\left(12\,C_{1}^{2}+18\,\mathrm{r_7}^{2}\right)-24\,\mathrm{r_3}^{2}\,\left(C_{1}^{2}+2\,\mathrm{r_7}^{2}\right)-6\,\sqrt{3}\,\mathrm{r_2}\,\mathrm{r_4}\,\mathrm{r_7}\,C_{4}\right)\\
&&+\partial_{t}r_{7}\,\left(6\,\mathrm{r_7}\,\left(\mathrm{r_2}\,\left(\mathrm{r_2}\,\mathrm{r_4}-2\,\mathrm{r_3}\,\mathrm{r_7}\right)-2\,\sqrt{3}\,\mathrm{r_3}\,\mathrm{r_4}\,C_{4}\right)+12\,\mathrm{r_2}^{3}\,C_{1}+6\,\mathrm{r_2}\,C_{1}\,\left(4\,\mathrm{r_3}^{2}+\mathrm{r_4}^{2}\right)\right)\\
&&+\partial_{t}r_{8}\,\left(\left(3\,\mathrm{r_8}+3\,\sqrt{3}\,\mathrm{r_3}\right)\,\mathrm{r_4}^{3}+\left(3\,\mathrm{r_7}^{2}\,\left(\mathrm{r_8}+3\,\sqrt{3}\,\mathrm{r_3}\right)+12\,\mathrm{r_2}^{2}\,\mathrm{r_8}-6\,\mathrm{r_3}\,\left(\mathrm{r_8}+\sqrt{3}\,\mathrm{r_3}\right)\,\left(\mathrm{r_3}-\sqrt{3}\,\mathrm{r_8}\right)\right)\,\mathrm{r_4}\right)\\
&&+\partial_{t}r_{8}\,\left(\mathrm{r_2}\,\mathrm{r_7}\,\left(6\,\sqrt{3}\,\mathrm{r_2}^{2}+4\,\sqrt{3}\,\mathrm{r_3}^{2}+\sqrt{3}\,\mathrm{r_4}^{2}-3\,\sqrt{3}\,\mathrm{r_7}^{2}\right)+6\,\frac{\sqrt{3}}{3}\,\mathrm{r_2}\,\mathrm{r_7}\,C_{4}\,\left(\mathrm{r_8}+2\,C_{3}\right)\right)\\
&&+\Gamma\,\left(\mathrm{r_2}^{2}\,\left(24\,\mathrm{r_4}\,\mathrm{r_7}^{2}-6\,\mathrm{r_4}\,\left(\mathrm{r_8}-\sqrt{3}\,\mathrm{r_3}\right)\,\left(\sqrt{3}\,\mathrm{r_3}-5\,\mathrm{r_8}+1\right)\right)-6\,\mathrm{r_2}^{3}\,\mathrm{r_7}\,\left(\mathrm{r_3}+\sqrt{3}\,\mathrm{r_8}-\sqrt{3}\right)\right)\\
&&+\Gamma\,\left(\mathrm{r_2}\,\left(\left(6\,C_{2}+6\,\sqrt{3}\right)\,\mathrm{r_7}^{3}+\left(\mathrm{r_4}^{2}\,\left(8\,\sqrt{3}\,C_{4}+4\,\sqrt{3}\right)+4\,\sqrt{3}\,\left(\sqrt{3}\,C_{1}C_{4}+4\,\mathrm{r_3}^{2}\right)+4\,\sqrt{3}\,C_{2}^{2}\,C_{1}\right)\,\mathrm{r_7}\right)\right)\\
&&+\Gamma\,\left(\mathrm{r_4}\,\mathrm{r_7}^{2}\,\left(6\,C_{4}+\sqrt{3}\,C_{4}\,\left(18\,\mathrm{r_3}+2\,\sqrt{3}\,\mathrm{r_8}\right)\right)-6\,\mathrm{r_4}^{3}\,C_{4}\left(C_{4}+2\right)+12\,\mathrm{r_3}\,C_{1}\,C_{4}\,\mathrm{r_4}\,\left(C_{4}-1\right)\right),
\end{eqnarray*}
\begin{eqnarray*}
n_{-1}&=&\partial_{t}r_{2}\,\left(6\,\sqrt{3}\,\mathrm{r_4}\,\mathrm{r_7}\,\left(C_{1}^{2}-3\,\mathrm{r_2}^{2}+\mathrm{r_4}^{2}+2\,\mathrm{r_7}^{2}\right)-6\,\sqrt{3}\,\mathrm{r_2}^{3}\,C_{1}-\sqrt{3}\,\mathrm{r_2}\,\left(6\,C_{1}^{2}C_{2}-12\,\mathrm{r_4}^{2}\,C_{1}+12\,\mathrm{r_7}^{2}\,C_{2}\right)\right)\\
&&+\partial_{t}r_{3}\,\left(-3\,\sqrt{3}\,\mathrm{r_2}^{2}\,\left(\left(\mathrm{r_4}^{2}-\mathrm{r_7}^{2}\right)+6\,\mathrm{r_3}\,C_{1}\right)-\sqrt{3}\,\mathrm{r_7}^{2}\,\left(9\,\mathrm{r_3}+\left(\sqrt{3}+2\right)\,\mathrm{r_8}\right)\,\left(\mathrm{r_3}-\left(\sqrt{3}-2\right)\,\mathrm{r_8}\right)\right)\\
&&+\partial_{t}r_{3}\,\left(3\,\sqrt{3}\,\left(3\,\mathrm{r_7}^{4}-\mathrm{r_4}^{4}\right)-6\,\sqrt{3}\,\mathrm{r_3}\,C_{2}\,C_{1}^{2}+9\,\mathrm{r_4}^{2}\,C_{1}C_{3}-24\,\sqrt{3}\,\mathrm{r_2}\,\mathrm{r_3}\,\mathrm{r_4}\,\mathrm{r_7}\right)\\
&&+\partial_{t}r_{4}\,\left(6\,\sqrt{3}\,\mathrm{r_7}\,\mathrm{r_2}\,\left(C_{1}^{2}-3\,\mathrm{r_4}^{2}+2\,\mathrm{r_7}^{2}+\mathrm{r_2}^{2}\right)+\sqrt{3}\,\mathrm{r_4}\,\left(6\,C_{1}\left(\mathrm{r_4}^{2}-2\,\mathrm{r_2}^{2}\right)-12\,\mathrm{r_3}\,\left(C_{1}^{2}+2\,\mathrm{r_7}^{2}\right)\right)\right)\\
&&+\partial_{t}r_{7}\,\left(\sqrt{3}\,\left(6\,\mathrm{r_7}\,\left(2\,\sqrt{3}\,\mathrm{r_8}\,\mathrm{r_2}^{2}-\sqrt{3}\,\mathrm{r_4}^{2}C_{4}\right)+6\,\mathrm{r_2}\,\mathrm{r_4}\,\left(C_{1}^{2}+\mathrm{r_2}^{2}+\mathrm{r_4}^{2}-2\,\mathrm{r_7}^{2}\right)\right)\right)\\
&&+\partial_{t}r_{8}\,\left(-\mathrm{r_2}^{2}\,\left(-\mathrm{r_2}^{2}+3\,\mathrm{r_4}^{2}+9\,\mathrm{r_7}^{2}+6\,C_{1}\,\left(\mathrm{r_3}+C_{2}\right)\right)-6\,\mathrm{r_3}\,C_{2}\,C_{1}^{2}\right)\\
&&+\partial_{t}r_{8}\,\left(3\,\mathrm{r_4}^{2}\,\left(\mathrm{r_4}^{2}+3\,\mathrm{r_7}^{2}-C_{1}^{2}\right)+6\,\mathrm{r_7}^{4}-\mathrm{r_7}^{2}\,\left(9\,\mathrm{r_3}+\left(\sqrt{3}+2\right)\,\mathrm{r_8}\right)\,\left(\mathrm{r_3}-\left(\sqrt{3}-2\right)\,\mathrm{r_8}\right)-24\,\sqrt{3}\,\mathrm{r_2}\,\mathrm{r_4}\,\mathrm{r_7}\,\mathrm{r_8}\right)\\
&&+\Gamma\,\left(\sqrt{3}\,\mathrm{r_2}^{4}\,\left(2\,\sqrt{3}\,\left(2\,\mathrm{r_8}-1\right)-6\,C_{1}\right)-24\,\mathrm{r_2}\,\mathrm{r_4}\,\mathrm{r_7}\,\left(\sqrt{3}\,\left(\mathrm{r_2}^{2}-\mathrm{r_4}^{2}\right)-\sqrt{3}\,C_{4}+2\,\sqrt{3}\,\left(\mathrm{r_3}^{2}-\mathrm{r_8}^{2}\right)\right)\right)\\
&&+\Gamma\,\left(-\mathrm{r_2}^{2}\,\left(6\,C_{1}^{2}\,\left(3\,C_{3}+2\,\mathrm{r_8}-1\right)+\sqrt{3}\,\mathrm{r_4}^{2}\,\left(2\,\sqrt{3}\,\left(2\,\mathrm{r_8}-1\right)-18\,C_{1}\right)+6\,\sqrt{3}\,\mathrm{r_7}^{2}\,\left(C_{2}+\sqrt{3}\right)\right)\right)\\
&&+\Gamma\,\left(12\,\left(1-C_{1}\right)\,\mathrm{r_4}^{4}+\left(18\,\mathrm{r_7}^{2}\,\left(C_{4}+1\right)-2\,\sqrt{3}\,C_{1}\,\left(4\,\mathrm{r_3}+C_{2}\right)\,\left(-2\,C_{1}-\sqrt{3}\,\left(\mathrm{r_8}-1\right)\right)\right)\,\mathrm{r_4}^{2}\right)\\
&&+\Gamma\,\left(12\,\mathrm{r_3}\,C_{2}\,C_{1}^{2}-12\,\mathrm{r_7}^{4}+18\,\mathrm{r_7}^{2}\,\left(\mathrm{r_3}+\left(\sqrt{3}+2\right)\,\mathrm{r_8}\right)\,\left(\mathrm{r_3}-\left(\sqrt{3}-2\right)\,\mathrm{r_8}\right)\right)\,\left(1-C_{4}\right),
\end{eqnarray*}
\begin{eqnarray*}
n_{+1}&=&\partial_{t}r_{2}\,\left(6\,\mathrm{r_2}\,\left(\mathrm{r_7}^{2}\,C_{3}-\mathrm{r_4}^{2}\,C_{4}\right)-2\,\sqrt{3}\,\mathrm{r_4}\,\mathrm{r_7}\,\left(-2\,\mathrm{r_2}^{2}+4\,\mathrm{r_3}^{2}+\mathrm{r_4}^{2}+\mathrm{r_7}^{2}\right)\right)\\
&&+\partial_{t}r_{3}\,\left(\left(\mathrm{r_4}^{2}+\mathrm{r_7}^{2}\right)\,\left(6\,\mathrm{r_3}\,\mathrm{r_8}+\sqrt{3}\,\left(\mathrm{r_4}^{2}-\mathrm{r_7}^{2}\right)\right)+2\,\sqrt{3}\,\left(\mathrm{r_4}^{2}\,\left(\mathrm{r_2}+\mathrm{r_3}\right)^{2}-\mathrm{r_7}^{2}\,\left(\mathrm{r_2}-\mathrm{r_3}\right)^{2}\right)\right)\\
&&+\partial_{t}r_{4}\,\left(4\,\sqrt{3}\,\left(\mathrm{r_2}^{2}+2\,\mathrm{r_3}^{2}\right)\,\mathrm{r_4}\,C_{1}-4\,\sqrt{3}\,\mathrm{r_3}\,\mathrm{r_4}\,\left(\mathrm{r_4}^{2}-2\,\mathrm{r_7}^{2}\right)-2\,\sqrt{3}\,\mathrm{r_2}\,\mathrm{r_7}\,\left(2\,\mathrm{r_2}^{2}+4\,\mathrm{r_3}^{2}-3\,\mathrm{r_4}^{2}+\mathrm{r_7}^{2}\right)\right)\\
&&+\partial_{t}r_{7}\,\left(-4\,\sqrt{3}\,\mathrm{r_2}^{2}\,\mathrm{r_7}\,C_{2}-4\,\sqrt{3}\,\mathrm{r_3}\,\mathrm{r_7}\,\left(2\,\mathrm{r_4}^{2}-\mathrm{r_7}^{2}+2\,\mathrm{r_3}\,C_{2}\right)-2\,\sqrt{3}\,\mathrm{r_2}\,\mathrm{r_4}\,\left(2\,\mathrm{r_2}^{2}+4\,\mathrm{r_3}^{2}+\mathrm{r_4}^{2}-3\,\mathrm{r_7}^{2}\right)\right)\\
&&+\partial_{t}r_{8}\,\left(4\,\left(\mathrm{r_2}^{2}+C_{1}C_{2}\right)\,\left(\mathrm{r_2}^{2}+2\,\mathrm{r_3}^{2}\right)-\left(2\,\mathrm{r_3}^{2}+\mathrm{r_4}^{2}+\mathrm{r_7}^{2}\right)\,\left(\mathrm{r_4}^{2}+\mathrm{r_7}^{2}\right)+8\,\sqrt{3}\,\mathrm{r_2}\,\mathrm{r_4}\,\mathrm{r_7}\,\mathrm{r_8}-6\,\sqrt{3}\,\mathrm{r_3}\,\mathrm{r_8}\,\left(\mathrm{r_4}^{2}-\mathrm{r_7}^{2}\right)\right)\\
&&+\Gamma\,\left(\left(2\,\mathrm{r_8}+6\,\sqrt{3}\,\mathrm{r_3}+2\right)\,\mathrm{r_7}^{4}+4\,\left(\mathrm{r_2}^{4}-\mathrm{r_4}^{4}\right)\,\left(1-C_{4}\right)\right)\\
&&+\Gamma\,\left(-\mathrm{r_2}^{2}\,\left(6\,\mathrm{r_4}^{2}\,\left(C_{4}+1\right)+2\,\mathrm{r_7}^{2}\,\left(\sqrt{3}\,C_{2}-3\right)-12\,\left(\mathrm{r_3}^{2}-\mathrm{r_8}^{2}\right)\,(1-C_{4})\right)\right)\\
&&+\Gamma\,\left(4\,\mathrm{r_3}\,\mathrm{r_4}^{2}\,\left(\mathrm{r_3}-3\,\sqrt{3}\,\mathrm{r_8}\right)\,\left(1-C_{4}\right)-8\,\sqrt{3}\,\mathrm{r_2}\,\mathrm{r_4}\,\mathrm{r_7}\,\left(\mathrm{r_3}^{2}-\mathrm{r_8}^{2}+\mathrm{r_4}^{2}-\mathrm{r_7}^{2}+2\,\sqrt{3}\,\mathrm{r_3}\,\mathrm{r_8}+C_{4}\right)\right)\\
&&+\Gamma\,\left(8\,\mathrm{r_3}^{2}\,C_{1}C_{2}\,\left(1-C_{4}\right)-\mathrm{r_7}^{2}\,\left(2\,\mathrm{r_4}^{2}\,\left(\mathrm{r_8}+7\,\sqrt{3}\,\mathrm{r_3}+1\right)+8\,\mathrm{r_3}^{2}\,\left(7\,\mathrm{r_8}+\sqrt{3}\,\mathrm{r_3}+1\right)\right)\right),
\end{eqnarray*}
with $C_{1}=\mathrm{r_{3}}-\sqrt{3}\,\mathrm{r_{8}},\,C_{2}=\mathrm{r_{3}}+\sqrt{3}\,\mathrm{r_{8}},\,C_{3}=\sqrt{3}\,
\mathrm{r_{3}}+\mathrm{r_8},\,C_{4}=\sqrt{3}\,\mathrm{r_3}-\mathrm{r_8}.$
\end{widetext}

\section{The Bloch vectors of the Trajectory  without Initial-to-Final State Couplings}\label{AE}

To obtain  a reasonable control parameters,  the intermediate state $\ket{^1E}$ is allowed to be occupied. Without loss of generality, the trajectory charactered by the Bloch vector is designed as follows:
\begin{eqnarray*}
r_3&=&\sqrt{3}\,\left(\left(2\,\mathrm{N}\,\left(3\,\mathrm{N}+2\right)\,\mathrm{\Gamma}^{2}+\mathrm{\Omega_{p}}^{2}-6\,\mathrm{\Omega_{p}}\,\mathrm{\Omega_{s}}+\mathrm{\Omega_{s}}^{2}\right)\right.\nonumber\\
&&+\cos\!\left(2\phi(t)\right)\,\left(3\,\left(\mathrm{\Omega_{p}}+\mathrm{\Omega_{s}}\right)^{2}+6\,\mathrm{N}\,\mathrm{\Gamma}^{2}\,\left(3\,\mathrm{N}+2\right)\right).\nonumber\\
&&-4\,\cos\!\left(\phi(t)\right)\,\left(\mathrm{\Omega_{p}}^{2}-\mathrm{\Omega_{s}}^{2}\right)\nonumber\\
&&+4\,\sqrt{2}\,\mathrm{N}\,\mathrm{\Gamma}\,\sin\!\left(\phi(t)\right)\,\left(\mathrm{\Omega_{p}}+\mathrm{\Omega_{s}}\right)\nonumber\\
&&\left.+6\,\sqrt{2}\,\mathrm{N}\,\mathrm{\Gamma}\,\sin\!\left(2\,\phi(t)\right)\,\left(\mathrm{\Omega_{p}}-\mathrm{\Omega_{s}}\right)\right)/(16z).\\
\end{eqnarray*}
\begin{eqnarray*}
r_8&=&\left(-\left(2\,\mathrm{N}\,\left(3\,\mathrm{N}+2\right)\,\mathrm{\Gamma}^{2}+\mathrm{\Omega_{p}}^{2}-6\,\mathrm{\Omega_{p}}\,\mathrm{\Omega_{s}}+\mathrm{\Omega_{s}}^{2}\right)\right.\nonumber\\
&&-12\,\cos\!\left(\phi(t)\right)\,\left(\mathrm{\Omega_{p}}^{2}-\mathrm{\Omega_{s}}^{2}\right)\nonumber\\
&&-\cos\!\left(2\,\phi(t)\right)\,\left(3\,\left(\mathrm{\Omega_{p}}+\mathrm{\Omega_{s}}\right)^{2}+6\,\mathrm{N}\,\mathrm{\Gamma}^{2}\,\left(3\,\mathrm{N}+2\right)\right)\nonumber\\
&&+12\,\sqrt{2}\,\mathrm{N}\,\mathrm{\Gamma}\,\sin\!\left(\phi(t)\right)\,\left(\mathrm{\Omega_{p}}+\mathrm{\Omega_{s}}\right)\nonumber\\
&&\left.-6\,\sqrt{2}\,\mathrm{N}\,\mathrm{\Gamma}\,\sin\!\left(2\,\phi(t)\right)\,\left(\mathrm{\Omega_{p}}-\mathrm{\Omega_{s}}\right)\right)/(16z),
\end{eqnarray*}
in which $\phi(t)$ determines the population on $\ket{^1E}$, i.e.,
\begin{eqnarray}
P_2&=&\left(2\, \mathrm{N}\, {\mathrm{\Omega}}^2 + 2\, {\mathrm{N}}^2\, {\mathrm{\Gamma}}^2\, \left(3\, \mathrm{N} + 2\right)\right.\nonumber\\
&& + {\sin\!\left(\phi(t)\right)}^2\, \left({\left(\mathrm{\Omega_p} + \mathrm{\Omega_s}\right)}^2 + 2\, \mathrm{N}\, {\mathrm{\Gamma}}^2\, \left(3\, \mathrm{N} + 2\right)\right) \nonumber\\
&&\left. -\sqrt{2}\, \mathrm{N}\, \mathrm{\Gamma}\, \sin\!\left(2\,\phi(t)\right)\, \left( \mathrm{\Omega_p} - \mathrm{\Omega_s}\right)\right)/(2z).
\end{eqnarray}
$r_4$ is determined by the implicit differential equation $\Omega_c^i(t)=0$; $r_2$ and $r_7$  are the same as in the adiabatic trajectory ( $r_2$ and $r_7$ in Eq.(\ref{r38})).
We consider the adiabatic pulses $\Omega_{p,s}$ like Eq.(\ref{ap3}) with $$\theta(t)=\frac{\pi}{2}\sin\!\left(\frac{\pi t}{2\tau}\right)^2,$$ and the trajectory with $$\phi(t)=\frac{\pi}{9}\sin\!\left(\frac{\pi t}{\tau}\right)^2 .$$
Thus, we can determine all of the control parameters numerically.

\end{document}